\documentclass[twocolumn,trackchanges,twocolappendix]{aastex631}
\shorttitle{}
\graphicspath{{./}{figures}}

\usepackage{enumitem}
\usepackage{amsmath}
\usepackage{tablefootnote}
\usepackage{longtable}

\begin{document}

\title{An Archival Optical Counterpart Search for Extragalactic Fast X-Ray Transients Discovered by Einstein Probe}
%Proof: TRT Thailand, ack, software
\author[0000-0001-6223-840X]{Run-Duo~Liang}
\affiliation{National Astronomical Observatories, Chinese Academy of Sciences, Beijing 100101, China}
\affiliation{University of Chinese Academy of Sciences, Chinese Academy of Sciences, Beijing 100049, China}
%\email{liangrd@bao.ac.cn}

\author[0000-0002-0096-3523]{Wen-Xiong~Li}
\correspondingauthor{Wen-Xiong~Li}
\email{liwx@bao.ac.cn}
\affiliation{National Astronomical Observatories, Chinese Academy of Sciences, Beijing 100101, China}

\author[0000-0002-8708-0597]{Liang-Duan~Liu}
\affiliation{Institute of Astrophysics, Central China Normal University, Wuhan, China}
\affiliation{Key Laboratory of Quark and Lepton Physics (Central China Normal University), Ministry of Education, Wuhan, China}

\author[0000-0001-9535-3199]{Ken W. Smith}
\affiliation{Astrophysics sub-Department, Department of Physics, University of Oxford, Keble Road, Oxford, OX1 3RH, UK}
\affiliation{Astrophysics Research Centre, School of Mathematics and Physics, Queen’s University Belfast, BT7 1NN, UK}

\author[0000-0002-8229-1731]{Stephen J. Smartt}
\affiliation{Astrophysics sub-Department, Department of Physics, University of Oxford, Keble Road, Oxford, OX1 3RH, UK}
\affiliation{Astrophysics Research Centre, School of Mathematics and Physics, Queen’s University Belfast, BT7 1NN, UK}

\author[0009-0002-9275-715X]{Qin-Yu~Wu}
\affiliation{National Astronomical Observatories, Chinese Academy of Sciences, Beijing 100101, China}
\affiliation{University of Chinese Academy of Sciences, Chinese Academy of Sciences, Beijing 100049, China}

\author[0000-0002-0656-075X]{Niu~Li}
\affiliation{National Astronomical Observatories, Chinese Academy of Sciences, Beijing 100101, China}

\author{Arne~Rau}
\affiliation{Max Planck Institute for extraterrestrial Physics, Giessenbachstr. 1, 85748 Garching, Germany}

\author[0000-0002-1094-3817]{Ling-Zhi~Wang}
\affiliation{Chinese Academy of Sciences South America Center for Astronomy (CASSACA), National Astronomical Observatories, CAS, Beijing 100101, China}
\affiliation{Departamento de Astronom{\'i}a, Universidad de Chile, Las Condes, 7591245, Santiago, Chile}

\author[0000-0002-4410-5387]{Armin Rest}
\affil{Space Telescope Science Institute, 3700 San Martin Dr., Baltimore, MD 21218, USA}
\affil{Physics and Astronomy Department, Johns Hopkins University, Baltimore, MD 21218, USA}

\author[0000-0002-4731-9698]{Ning-Chen Sun}
\affiliation{National Astronomical Observatories, Chinese Academy of Sciences, Beijing 100101, China}
\affiliation{School of Astronomy and Space Science, University of Chinese Academy of Sciences, Beijing, 100049, China}
\affiliation{Institute for Frontiers in Astronomy and Astrophysics, Beijing Normal University, Beijing, 102206, China}

\author[0000-0002-8686-8737]{Franz E. Bauer}
\affiliation{Instituto de Alta Investigaci{\'o}n, Universidad de Tarapac{\'a}, Casilla 7D, Arica, Chile}

\author[0000-0001-7568-6412]{Ezequiel Treister}
\affiliation{Instituto de Alta Investigaci{\'o}n, Universidad de Tarapac{\'a}, Casilla 7D, Arica, Chile}

\author[0000-0001-6511-8745]{Jia-Sheng Huang}
\affiliation{Chinese Academy of Sciences South America Center for Astronomy (CASSACA), National Astronomical Observatories, CAS, Beijing 100101, China}

\author[0009-0000-6374-3221]{Jennifer Chac{\'o}n}
\affiliation{Instituto de Astrofísica, Pontificia Universidad Cat{\'o}lica de Chile, Santiago, Chile}

\author[0000-0003-1325-6235]{Se\'an~J.~Brennan}
\affiliation{Max Planck Institute for extraterrestrial Physics, Giessenbachstr. 1, 85748 Garching, Germany}

\author[0000-0002-2555-3192]{Matt Nicholl}
\affiliation{Astrophysics Research Centre, School of Mathematics and Physics, Queens University Belfast, Belfast BT7 1NN, UK}

\author[0000-0002-1066-6098]{Ting-Wan~Chen}
\affiliation{Graduate Institute of Astronomy, National Central University, 300 Jhongda Road, 32001 Jhongli, Taiwan}

\author[0000-0002-9928-0369]{Amar~Aryan}
\affiliation{Graduate Institute of Astronomy, National Central University, 300 Jhongda Road, 32001 Jhongli, Taiwan}

\author[0000-0002-2898-6532]{Sheng~Yang}
\affiliation{Institute for Gravitational Wave Astronomy, Henan Academy of Sciences, Zhengzhou 450046, Henan, China}

\author[0000-0002-5105-344X]{Albert~K.H.~Kong}
\affiliation{Institute of Astronomy, National Tsing Hua University, Hsinchu 300044, Taiwan}

\author[0000-0002-3825-0553]{Sofia Rest}
\affil{Physics and Astronomy Department, Johns Hopkins University, Baltimore, MD 21218, USA}

\author[0000-0001-5233-6989]{Qi-Nan~Wang}
\affiliation{Department of Physics and Kavli Institute for Astrophysics and Space Research, Massachusetts Institute of Technology, 77 Massachusetts Avenue, Cambridge, MA 02139, USA}

\author[0000-0002-8094-6108]{James~H.~Gillanders}
\affiliation{Astrophysics sub-Department, Department of Physics, University of Oxford, Keble Road, Oxford, OX1 3RH, UK}

\author[0000-0002-4562-7179]{Dong-Yue~Li}
\affiliation{National Astronomical Observatories, Chinese Academy of Sciences, Beijing 100101, China}

\author[0000-0002-0823-4317]{An~Li}
\affiliation{Department of Astronomy, Beijing Normal University, Beijing 100875, People's Republic of China}
\affiliation{Institute for Frontier in Astronomy and Astrophysics, Beijing Normal University, Beijing 102206, People's Republic of China}

\author[0000-0002-5485-5042]{Jun~Yang}
\affiliation{Institute for Astrophysics, School of Physics, Zhengzhou University, Zhengzhou 450001, China}

\author[0000-0001-9893-8248]{Qing-Chang Zhao}
\affiliation{University of Chinese Academy of Sciences, Chinese Academy of Sciences, Beijing 100049, China}
\affiliation{Key Laboratory of Particle Astrophysics, Institute of High Energy Physics, Chinese Academy of Sciences, Beijing 100049, China} 
%\email{zhaoqc@ihep.ac.cn}

\author[0000-0002-9615-1481]{Hui~Sun}
\affiliation{National Astronomical Observatories, Chinese Academy of Sciences, Beijing 100101, China}

%\author[0000-0002-6299-1263]{Xue-Feng~Wu}
%\affiliation{Purple Mountain Observatory, Chinese Academy of Sciences, Nanjing 210023, People’s Republic of China}

\author[0000-0002-7397-811X]{Yun-Fei~Xu}
\affiliation{National Astronomical Observatories, Chinese Academy of Sciences, Beijing 100101, China}
\affiliation{National Astronomical Data Centre of China, Beijing 100101, China}

\author[0009-0008-7068-0693]{Zhi-Xing~Ling}
\affiliation{National Astronomical Observatories, Chinese Academy of Sciences, Beijing 100101, China}
\affiliation{School of Astronomy and Space Science, University of Chinese Academy of Sciences, 19A Yuquan Road, Beijing 100049, People’s Republic of China}
\affiliation{Institute for Frontier in Astronomy and Astrophysics, Beijing Normal University, Beijing 102206, People’s Republic of China}

\author[0000-0001-5486-2747]{Thomas J. L. de Boer} % Pan-STARRS co-author
\affiliation{Institute for Astronomy, University of Hawai`i, 2680 Woodlawn Drive, Honolulu HI 96822, USA}

\author[0000-0001-6965-7789]{Ken~C.~Chambers} % Pan-STARRS co-author
\affiliation{Institute for Astronomy, University of Hawai`i, 2680 Woodlawn Drive, Honolulu HI 96822, USA}

\author[0000-0002-7272-5129]{Chien-Cheng Lin} % Pan-STARRS co-author
\affiliation{Institute for Astronomy, University of Hawai`i, 2680 Woodlawn Drive, Honolulu HI 96822, USA}

\author[0000-0002-9438-3617]{Thomas B. Lowe} % Pan-STARRS co-author
\affiliation{Institute for Astronomy, University of Hawai`i, 2680 Woodlawn Drive, Honolulu HI 96822, USA}

\author[0000-0002-7965-2815]{Eugene~A.~Magnier} % Pan-STARRS co-author
\affiliation{Institute for Astronomy, University of Hawai`i, 2680 Woodlawn Drive, Honolulu HI 96822, USA}

\author[0000-0002-1341-0952]{Richard J.~Wainscoat} % Pan-STARRS co-author
\affiliation{Institute for Astronomy, University of Hawai`i, 2680 Woodlawn Drive, Honolulu HI 96822, USA}

\author[0000-0001-8602-4641]{J.~Quirola-V{\'a}squez}
\affiliation{Department of Astrophysics/IMAPP, Radboud University, PO Box 9010, 6500 GL, The Netherlands}

\author{Xiao-Feng~Wang}
\affiliation{Department of Physics, Tsinghua University, Beijing, 100084, China}

\author[0000-0002-1481-4676]{Samaporn~Tinyanont}
\affiliation{National Astronomical Research Institute of Thailand, 260 Moo 4, Donkaew, Maerim, Chiang Mai, 50180, Thailand}
%\email{samaporn@narit.or.th}

\author[0000-0002-0779-1947]{Jing-Wei~Hu}
\affiliation{National Astronomical Observatories, Chinese Academy of Sciences, Beijing 100101, China}

\author[0000-0002-2412-5751]{He-Yang~Liu}
\affiliation{National Astronomical Observatories, Chinese Academy of Sciences, Beijing 100101, China}

\author[0000-0003-4200-9954]{Hua-Qing~Cheng}
\affiliation{National Astronomical Observatories, Chinese Academy of Sciences, Beijing 100101, China}

\author{Hao-Wei~Peng}
\affiliation{Department of Physics, Tsinghua University, Beijing, 100084, China}

\author{Chen~Zhang}
\affiliation{National Astronomical Observatories, Chinese Academy of Sciences, Beijing 100101, China}

\author{Dong-Hua~Zhao}
\affiliation{National Astronomical Observatories, Chinese Academy of Sciences, Beijing 100101, China}

\author{Mao-Hai~Huang}
\affiliation{National Astronomical Observatories, Chinese Academy of Sciences, Beijing 100101, China}

\author[0000-0001-9834-2196]{Yong~Chen}
\affiliation{Key Laboratory of Particle Astrophysics, Institute of High Energy Physics, Chinese Academy of Sciences, Beijing 100049, China}

\author[0000-0002-5203-8321]{Shu-Mei~Jia}
\affiliation{Key Laboratory of Particle Astrophysics, Institute of High Energy Physics, Chinese Academy of Sciences, Beijing 100049, China}

\author[0000-0001-5798-4491]{Cheng-Kui~Li}
\affiliation{Key Laboratory of Particle Astrophysics, Institute of High Energy Physics, Chinese Academy of Sciences, Beijing 100049, China}

\author{Ju~Guan}
\affiliation{Key Laboratory of Particle Astrophysics, Institute of High Energy Physics, Chinese Academy of Sciences, Beijing 100049, China}

\author{Chen-Zhou~Cui}
\affiliation{National Astronomical Observatories, Chinese Academy of Sciences, Beijing 100101, China}

\author[0009-0002-9275-715X]{Yuan~Liu}
\affiliation{National Astronomical Observatories, Chinese Academy of Sciences, Beijing 100101, China}

\author[0000-0001-8266-3024]{Weimin~Yuan}
\correspondingauthor{Weimin Yuan}
\email{wmy@nao.cas.cn}

\affiliation{National Astronomical Observatories, Chinese Academy of Sciences, Beijing 100101, China}
\affiliation{University of Chinese Academy of Sciences, Chinese Academy of Sciences, Beijing 100049, China}

\begin{abstract}

Extragalactic fast X-ray transients (eFXTs) represent a rapidly growing class of high-energy phenomena, whose physical origins remain poorly understood.  With its wide-field, sensitive all-sky monitoring, the \textit{Einstein Probe} (EP) has greatly increased the discovery rate of eFXTs. The search and identification of the optical counterparts of eFXT are vital for understanding their classification and constraining their physical origin.
Yet, a considerable fraction of eFXTs still lack secure classifications due to the absence of timely follow-up observations.
We carry out a systematic search of publicly available optical survey data and transient databases 
(including the Zwicky Transient Facility, ZTF, and the Transient Name Server, TNS) 
for optical counterparts to eFXT candidates detected by EP. In this paper, we describe our ongoing program and report the first results. Specifically, we identified the eFXT EP240506a to be associated with a UV/optical counterpart, AT\,2024ofs. Spectroscopy of its host galaxy with VLT yields a redshift of $z = 0.120 \pm 0.002$. By combining archival survey data with early-time multiwavelength observations, we find that the luminosity and light-curve evolution of AT~2024ofs are consistent with a core-collapse supernova origin. From detectability simulations, we estimate a local event rate density $\rho_{0}=8.8^{+21.2}_{-3.9}\ \mathrm{yr^{-1}\,Gpc^{-3}}$ for EP240506a-like events, and completeness-corrected rate of about $36$--$78\ \mathrm{yr^{-1}\ Gpc^{-3}}$ for EP-detected X-ray transients associated with supernovae. Our results demonstrate the potential of EP to uncover prompt high-energy emission from core-collapse supernovae and underscore the critical importance of timely follow-up of future eFXT events.

%The nature of extragalactic Fast X-ray Transients (eFXTs) is still elusive. The launch of \textit{Einstein Probe} (EP) has firmly promoted the discovery of X-ray transients with high-cadence scanning of the sky. EP has discovered a significant number of eFXTs, of which the origin of a fraction remains uncertain, due to the lack of timely Target-of-Opportunity (ToO) observations and spectroscopic identifications. In this work, we present a systematic search for optical counterparts of eFXT candidates detected by EP. An exception (surprise), EP240506a, was re-identified to be associated with a supernova candidate AT~2024ofs, as it was missed by initial ToO observations. We obtained the spectrum of the host galaxy with VLT, measuring the redshift $z=0.120\ (\pm 0.002)$. By digging out archival survey data and analysing the initial multi-wavelength data, the luminosity and the evolution support a core-collapse origin, with an estimated local event rate density $\rho_{0}=8.8^{+21.2}_{-3.9}\ \mathrm{yr^{-1}\ Gpc^{-3}}$. And the local event rate for X-ray transients associated with a SN is $\rho_{0} = 11.4^{+9.2}_{-2.2}\ \mathrm{yr^{-1}\ Gpc^{-3}}$. In addition, the authencity of other optical counterpart candidates. Our results highlight the potential of discovering prompt emission from core-collapse supernova for EP, and underscore the importance of timely follow-ups and searching. 

\end{abstract}


\section{Introduction} \label{sec:intro}

Over the past few decades, advances in wide-field and/or high-sensitivity facilities across the electromagnetic spectrum—from gamma-rays to radio—have progressively unveiled the dynamic nature of celestial objects. Some distinct types of transients, including gamma-ray bursts (GRBs) and supernovae (SNe), have been systematically characterized. More recently, newly discovered phenomena such as fast radio bursts (FRBs), fast blue optical transients (FBOTs), and kilonovae (KNe) have further expanded the diversity of known astrophysical transients~\citep{1983bhwd.book.....S,2019LRR....23....1M,zhang2019physics}. 

In the soft X-ray regime ($0.3-10.0$ keV), extragalactic fast X-ray transients (eFXTs) represent a class of short-duration X-ray flashes originating at cosmological distances, typically lasting from seconds to several hours. 
Early manifestations of such events were detected by X-ray monitors onboard missions such as \textit{BeppoSAX} and \textit{HETE-II}, some of which lacked clear gamma-ray counterparts and were referred as X-ray flashes~\citep[XRFs;][]{1997Natur.387..783C,lamb2004scientific}. 
On the other hand, eFXTs had also been identified through search of archival data obtained by narrow-field but highly sensitive X-ray telescopes such as \textit{Chandra}~\citep{weisskopf2000chandra,yang2019searching,quirola2022extragalactic,quirola2023extragalactic}, \textit{XMM-Newton}~\citealt{jansen2001xmm,alp2020blasts}), and the \textit{Neil Gehrels Swift} Observatory (\textit{Swift}-XRT;~\citealt{burrows2005swift,campana2006association,soderberg2008erratum,mazzali2008metamorphosis}~\citeyear{burrows2005swift}). 
Prior to early 2024, only several dozen eFXTs have been known, and their origin remains shrouded in uncertainty due to the lack of timely follow-up observations~\citep {yang2019searching,alp2020blasts,quirola2022extragalactic,quirola2023extragalactic}.

Shock breakout (SBO) from core-collapse SNe has long been considered a promising process for producing eFXTs.
Two well-established cases in observations are XRO~080109/SN~2008D and the relativistic SBO associated with GRB~060218/SN~2006aj, providing direct evidence that fast X-ray flashes can arise when the shock front emerges from the stellar envelope or from a dense circumstellar medium \citep{campana2006association,soderberg2008erratum,mazzali2008metamorphosis,nakar2010early,jonker2013discovery,waxman2017shock,novara2020supernova,alp2020blasts,deng2018detection,li2024shock}. Rather than representing distinct classes, these events are thought to be likely part of a continuous zoo of collapsar-related transients, shaped by variations in jet power, viewing angle, progenitor structure, and circumstellar environment. Within this framework, eFXTs may naturally bridge the most luminous long GRBs, softer or off-axis GRBs, and low-luminosity X-ray flashes~\citep{zhang2004gamma,nakar2015unified,d2018grb,zhang2019physics,liu2025soft,2025A&A...695A.279Q}. Additional pathways may also contribute to the observed eFXT population, including mildly relativistic cocoons or central-engine-powered X-ray emission following compact-object mergers \citep{zhang2013early,xue2019magnetar,metzger2020kilonovae,sun2025magnetar}, and rare tidal disruption events involving white dwarfs and intermediate-mass black holes \citep{jonker2013discovery,glennie2015two}. 

%Shock breakout (SBO) of core-collapse SNe is proposed to explain some eFXTs, supported by two well-established cases: the discovery of XRO 080109 associated with SN 2008D, and the relativistic SBO observed in GRB 060218 associated with SN 2006aj~\citep{campana2006association,soderberg2008erratum,mazzali2008metamorphosis}. These events provide strong observational evidence that eFXTs can arise from the brief, luminous burst of high-energy radiation produced when the SN shock front emerges from the stellar envelope or a dense circumstellar medium~\citep{nakar2010early,waxman2017shock,novara2020supernova,alp2020blasts,deng2018detection,li2024shock}. Several other mechanisms have been proposed to account for the origin of some eFXTs. \textcolor{red}{These include %softer or off-axis long GRBs (LGRBs)}~\citep{zhang2004gamma,nakar2015unified,d2018grb,zhang2019physics,liu2025soft}, mildly relativistic cocoons or X-ray emission powered by a long-lived central engine following binary neutron star mergers~\citep{jonker2013discovery,xue2019magnetar,metzger2020kilonovae,sun2025magnetar}, as well as tidal disruption events (TDEs) involving the disruption of a white dwarf by an intermediate-mass black hole~\citep{jonker2013discovery,glennie2015two}. %Recent observations further highlight the diversity of eFXTs, emphasizing that their physical origins remain only partially understood and are likely to involve multiple progenitor channels~\citep{zhang2025einstein,xinwen2025ep241021a}.

This situation has undergone a dramatic shift with the launch of the \textit{Einstein Probe} (\citealt{yuan2022einstein,yuan2025science}; EP) in 2024, equipped with an exceptionally wide field of view ($\sim 3600\ \mathrm{deg}^{2}$) and the capability for rapid follow-up observations through the Wide-field X-ray Telescope (WXT) and the Follow-up X-ray Telescope (FXT). In its first 1.5 years of operation, EP has discovered over 90 eFXTs\footnote{The final eFXT catalog of Wu et al. contains fewer sources than earlier GCN/ATel reports, as some candidates were later identified as other source classes or deemed non-genuine.}, a fraction of which remain of uncertain origin (Wu et al. in prep). Among these, five eFXTs (EP240414a, EP240801a, EP250108a, EP250304a and EP250827b) have been found to be associated with broad-lined type Ic (Ic-BL) SNe, opening a new window into unveiling the physical processes governing the death of massive stars~\citep{sun2025fast,hamidani2025ep240414a,2025ApJ...982L..47V,van2025einstein,li2025extremely,2025ApJ...988L..14E,eyles2025ep250108ahop,srinivasaragavan2025ep250108a,2025ApJ...988L..60S,Chen250304a}.

However, due to the unknown redshift in some cases, the delay to the rest-frame SN peak and the uncertain object faintness, SNe detection and confirmation may have been missed by inhomogeneous or limited follow-up efforts. Moreover, for sub-threshold events with large position uncertainty that lacked prompt multiwavelength follow-up, their optical counterparts may still be recovered serendipitously in archival or ongoing wide-field optical surveys. In this work, we conduct a systematic search for optical counterparts of EP-detected X-ray transients, aiming to probe the physical origins of these EP-discovered eFXTs.

The detectability of associated SNe varies substantially across the EP sample. For events with sub-$10^{\prime\prime}$ localizations, coordinated follow-up campaigns are typically extended for weeks, providing reasonably strong constraints on any underlying SN emission. In contrast, a large fraction of EP events lack precise localizations, leading to follow-up that was sparse, short-lived, or absent altogether. For these poorly localized or sub-threshold events, any accompanying SNe---especially those peaking several days after the X-ray trigger or intrinsically faint in the observer frame---could have been easily missed. Nonetheless, for events without prompt multiwavelength observations, their optical counterparts may still be recoverable serendipitously in archival or ongoing wide-field surveys. Motivated by these considerations, we conduct a systematic search for optical counterparts of EP-detected X-ray transients, aiming to probe the physical origins of the growing population of eFXTs discovered by EP.

The paper is organized as follows. In Section~\ref{sec:method}, we describe the selection criteria for the X-ray and optical samples and outline the cross-matching methodology. Section~\ref{sec:result} presents the cross-match results, with Section~\ref{subsec:Ep240506a} providing a detailed analysis of the newly identified X-ray transient EP240506a/AT~2024ofs. In Section~\ref{sec:discuss}, we discuss the probability of chance coincidence and estimate the event rate. Finally, we summarize our results in Section~\ref{sec:conclusion}. 

\section{Methodology and Data Selection}
\label{sec:method}

\subsection{X-ray Sample}

To ensure the completeness of our X-ray sample, we used a list of all high-confidence eFXT candidates (Wu et al. in prep) and \textit{unverified sources} detected by EP between January 9, 2024, and June 20, 2025. 

Generally, eFXTs are defined as non-Galactic sources that manifest as nonrepeating, soft X-ray flashes~\citep{alp2020blasts,quirola2023extragalactic}. To be included in our high-confidence eFXT candidate sample, a source must satisfy the following criteria based on the method described by Wu et al. (in prep):

\begin{enumerate}[label=(\alph*)]
    \item \textbf{High Significance}. The detection signal-to-noise ratio (S/N) must generally exceed 6.0 for ground-based triggers and 7.0 for on-board triggers. 
    \item \textbf{No X-ray History}. 
    %We cross-match the source localization against (i) the TA-verified EP/WXT internal source list and (ii) major public X-ray source catalogs (ROSAT, Swift/XRT, MAXI, and the MPE public eROSITA catalog). Candidates with a consistent cataloged X-ray counterpart are rejected. 
    There must be no known X-ray source previously detected at the position.
    \item \textbf{Not a Galactic origin}. 
    %We exclude candidates consistent with a Galactic object or transient. In practice, we cross-match the X-ray localization with archival optical images (e.g., DSS and Pan-STARRS) and optical catalogs (e.g., Gaia DR3) to identify likely stellar counterparts and to rule out scenarios such as stellar flares or Galactic X-ray binary activity; available multi-wavelength follow-up information is further used to reject candidates showing clear evidence for a Galactic origin. 
    The source must not be associated with any known Galactic object, e.g., X-ray binaries, stars. 
    \item \textbf{Short-duration X-ray flash}. The source must exhibit a distinct X-ray fluence enhancement confined within a single orbit observation (approximately 3.6\,ks). 
    %We apply the Bayesian Blocks algorithm to the single-observation light curve to test for the presence of variability, requiring at least one statistically significant change point in addition to those trivially associated with the observation start and end times.
\end{enumerate}

In addition to high-confidence eFXT candidates, we also include \textit{unverified sources} (sub-threshold sources) in our analysis\footnote{Their signal-to-noise ratios typically lie in the range between 5.0 and 6.0. The sub-threshold catalog will be published later.}. These events are labeled as \texttt{Unverified} during post-processing, often due to their low S/N, the absence of clear flaring behavior in the light curve, or the lack of prior X-ray detections. Such characteristics suggest that they are likely genuine detections, although their physical nature remains uncertain. In general, \textit{unverified sources} display realistic PSFs and reasonable spectral properties, but their classification is hindered by the relatively large WXT positional uncertainties, which stem from the limited data quality. We include them to avoid missing potential low-S/N eFXT candidates. 
%typically due to characteristics such as low signal-to-noise ratio (S/N), the absence of prominent flares in the light curve, or a lack of X-ray history detections, suggesting that they are genuine detections whose physical nature remains uncertain.  Generally, \textit{unverified sources} exhibit realistic PSFs and reasonable spectra, but it is challenging to identify their origin due to the large WXT positional uncertainties. This can be attributed to the limited data quality. Nonetheless, we include them to prevent potential low S/N eFXT candidates from being missing. 

After applying these filtering criteria, our selected X-ray sample comprises 95 eFXT candidates and 5,140 \textit{unverified sources}. The sky distribution can be seen in Figure~\ref{fig:sky}. 

\subsection{Optical Data Selection}

We constructed an initial sample of optical transients by applying a filtering process to the ZTF alert stream~\citep{patterson2018zwicky}, supplemented with objects reported in the Transient Name Server\footnote{https://www.wis-tns.org/} (TNS). To identify extragalactic optical transients, we utilized the Lasair broker system\footnote{https://lasair-ztf.lsst.ac.uk/}, which provides real-time access to ZTF alert data via structured SQL-based queries~\citep{2024RASTI...3..362W}. As part of this process, we leveraged the \texttt{Sherlock} sky context package, an integrated database framework that enables rapid and reliable classification by spatial cross-matching with a wide range of astrophysical catalogs, facilitating efficient selection of transients of interest~\citep{david_young_2023_8289325}. The optical candidates should satisfy the following criteria:

\begin{itemize}
    %\item The object appears on a designated watchlist (\texttt{wl\_id = 1}).
    \item The object is either cross-matched with a source in the TNS or has a \texttt{Sherlock} classification label as one of (\texttt{NT}, \texttt{SN}, \texttt{ORPHAN}). These labels correspond to the following categories, respectively: (1) transients that fall within the $1.5^{\prime\prime}$ of the core of a resolved galaxy (\texttt{NT}), (2) transients that are not classified as \texttt{NT} but are located within a magnitude-, morphology-, or distance-dependent association radius of a galaxy (\texttt{SN}), and (3) transients that fail to match any known catalogued source (\texttt{ORPHAN}). 
    \item The object is $>2^{\prime\prime}$ away from star-like objects ($\texttt{sgscore}>0.5$) in Pan-STARRS (PS) catalog~\citep{chambers2016pan}. 
    \item The discovery date reported by TNS is later than January 9, 2024 (UTC), corresponding to the launch date of EP.
    \item The earliest detection in ZTF is later than January 9, 2024 (UTC). 
\end{itemize}

\begin{figure}[htb]
    \centering
    \includegraphics[width=0.95\linewidth]{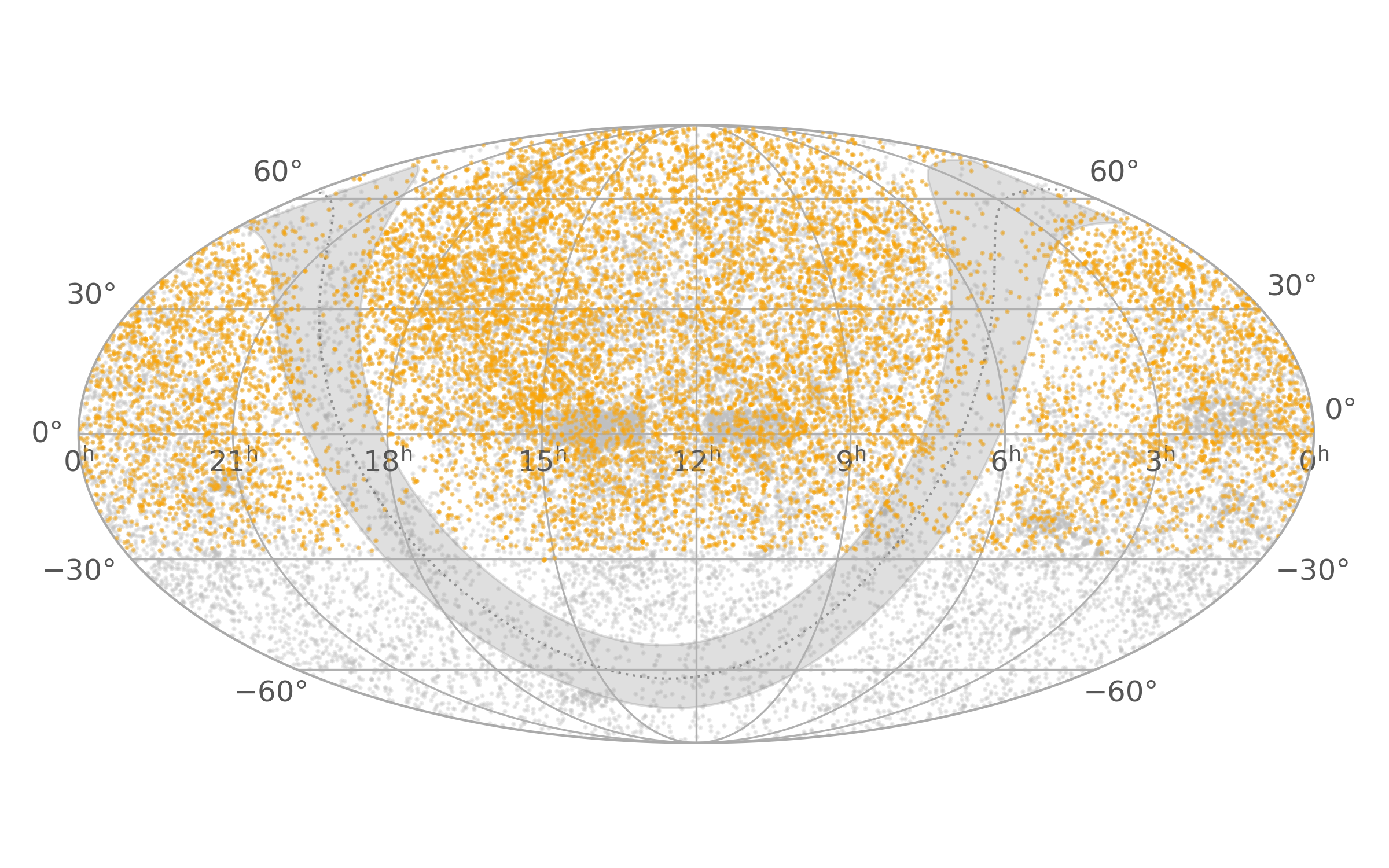}\\
    \includegraphics[width=0.95\linewidth]{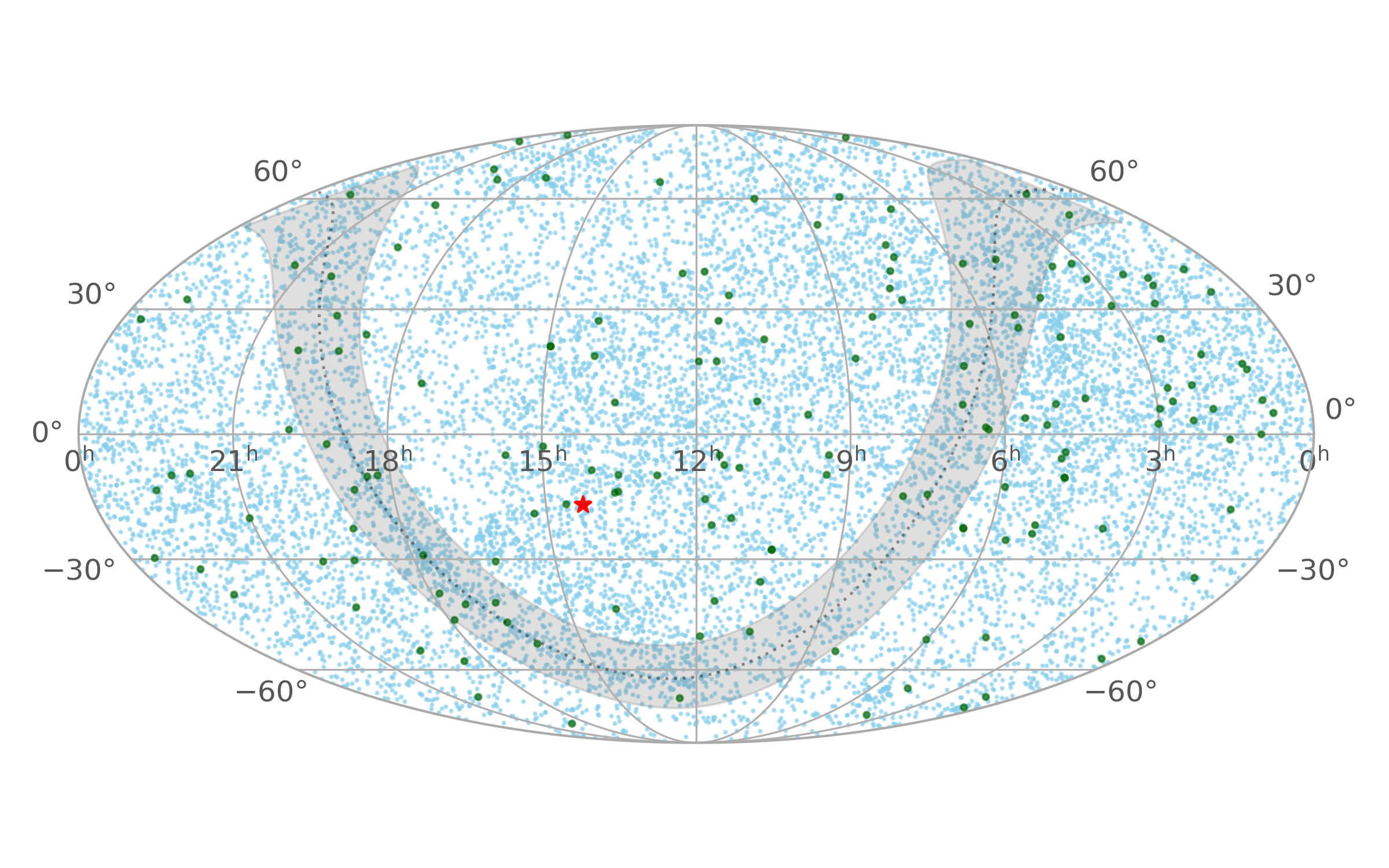}
    \caption{Sky distributions of the samples. The shaded region and the dotted line indicate Galactic latitudes $|b| < 10^\circ$ and the Galactic plane, respectively. \textit{Upper panel:} Optical sample, including TNS objects (grey dots) and ZTF transient candidates (orange dots). \textit{Lower panel:} \textit{unverified sources} (blue dots) and high-confidence eFXT candidates (green dots) detected by EP. The red star marks the position of EP240506a.}
    \label{fig:sky}
\end{figure}

It is important to note that EP commenced routine observations on March 1, 2024, following the gradual activation of all 48 CMOS modules. Therefore, the selected optical transient sample fully encompasses the temporal window during which associations with X-ray sources could plausibly occur.
This filtering strategy enabled the construction of a sample of extragalactic optical transients from the ZTF alert stream. After applying all selection criteria, we identified a total of 12,346 candidates, which form the basis for subsequent multi-wavelength cross-matching and analysis. In addition, we incorporated transients reported to the Transient Name Server (TNS) since the launch of EP, resulting in a supplementary sample of 33,048 objects. This sample partially overlaps with the ZTF-selected candidates, thereby improving the overall completeness of our transient catalog. The combined sky distribution of these samples is shown in Figure~\ref{fig:sky}. There is a concentrated region for \textit{unverified sources} at the Galactic plane between $15^\mathrm{h} -19^\mathrm{h}$, attributed to stellar flares and X-ray binaries. 

\subsection{Cross-matching of X-ray and Optical Transients}

With the compiled catalogs of X-ray and optical transients, we performed a cross-match to identify potential SN candidates exhibiting prompt X-ray emission. For each X-ray source, we identified optical transients residing within the positional uncertainty region of the WXT detection. A conservative matching radius of $r \le 3.5'$ was adopted to accommodate potential localization uncertainties, particularly given that the positional accuracy during the early commissioning phase of EP was approximately $3.5^{\prime}$~\citep{yuan2022einstein}. Subsequently, we applied a temporal constraint by selecting candidates within a time offset $\delta t$,  defined as the time difference between the EP detection and the optical discovery. A positive $\delta t$ indicates that the X-ray detection precedes the optical transient, aligning with the expected sequence for phenomena such as SN SBO. Considering the expected delay between the prompt X-ray emission and the emergence of optical SN signatures~\citep{soderberg2008extremely,campana2006association, forster2018delay,sun2025fast,li2025extremely}, as well as the survey cadence and occasional gaps in optical coverage—which can cause the discovery time to lag behind the true explosion time \(t_0\) by missing the early rise or peak—we adopt a conservative matching window of \(0 < \delta t \le 30\ \mathrm{days}\). This choice is motivated by the fact that core-collapse SNe typically show rest-frame rise times of \(t_{\mathrm{r}} \lesssim 20\) days~\citep{2016MNRAS.458.2973P,2019A&A...621A..71T}. For the maximum redshift (\(\sim 0.4\)) of ground-based EP SN discoveries, this corresponds to an observer-frame rise time of \(t_{\mathrm{r}} \lesssim 28\) days, well within our adopted window.
%Considering the expected delay between prompt X-ray emission and the optical detection of SNe~\citep{soderberg2008extremely,campana2006association,forster2018delay,sun2025fast,li2025extremely}, as well as the observational cadence and potential gaps in optical survey coverage---which may cause the discovery date to lag behind the true explosion time $t_0$ by missing the early rise or peak phase---we adopt a conservative matching window of $0 < \delta t \le 30\ \mathrm{days}$ given that core-collapse SNe (CCSNe) typically exhibit rest-frame rise times $t_{\mathrm{r}} \lesssim 20$ days~\citep{2016MNRAS.458.2973P,2019A&A...621A..71T} and the corresponding $t_{\mathrm{r}} \lesssim 28$ in observer frame based on the maximum redsfhit of $\sim 0.4$ for ground-discovered EP SN. 

It is important to note that the candidates identified through this process include not only genuine candidates but also known events (i.e., GRB afterglows), contaminations (i.e., Galactic sources), and chance coincidences. Therefore, a thorough vetting and multi-wavelength analysis are essential for assessing the nature of each candidate.

\section{Results}
\label{sec:result}

\subsection{Optical candidates to X-ray Sample}\label{subsec:candidates}

% =====================
% High-confidence sources
% =====================
\begin{deluxetable*}{cccccccc}[htb]
\small
\tablecaption{List of high-confidence eFXTs and identified optical counterpart candidates.}
\label{table:candidates_highconf}
\tablehead{
\colhead{EP Name} & \colhead{$T_{0}$} & \colhead{$S_{X}$} & \colhead{TNS Name} & \colhead{Classification} & \colhead{Separation} & \colhead{$\delta t$} & \colhead{Reference} \\ 
& & & & & [arcminutes] & [days] & 
}
\startdata
EP240315a & 2024-03-15 18:27:18 & 28.88 & AT~2024eju & GRB & 0.91 & 1.086 & \scriptsize \citealt{liu2025soft,2024arXiv240416350L} \\ 
EP240414a & 2024-04-14 09:49:10 & 11.23 & AT~2024gsa & SN & 1.54 & 0.125 & \scriptsize \citealt{sun2025fast,van2025einstein} \\ 
EP240425a & 2024-04-25 20:49:24 & 8.34 & AT~2024ocl & Unknown & 0.89 & 8.593 & \scriptsize Zhao et al. in prep \\ 
EP240506a & 2024-05-06 05:01:39 & 7.80 & AT~2024ofs & SN candidate & 2.95 & 7.719 & \scriptsize \citealt{li2024ep240506a} \\ 
EP241030a & 2024-10-30 06:33:18 & 9.01 & AT~2024zuk & GRB & 1.25 & 0.435 & \scriptsize \citealt{2024GCN.37997....1W} \\ 
EP250108a & 2025-01-08 12:47:35 & 6.27 & SN~2025kg & SN & 0.31 & 0.970 & \scriptsize \citealt{li2025ep250108a,2025ApJ...988L..14E} \\ 
EP250226a & 2025-02-26 06:35:16 & 7.74 & AT~2025dbz & GRB & 0.48 & 3.373 & \scriptsize \citealt{jiang2025ep250226a} \\ 
EP250304a & 2025-03-04 01:32:30 & 14.58 & AT~2025fhm & SN & 0.82 & 0.825 & \scriptsize \citealt{Chen250304a} \\ 
EP250427a & 2025-01-08 12:47:35 & 6.27 & AT~2025inn & GRB & 0.31 & 0.970 & \scriptsize \citealt{wang2025ep250427a} \\ 
\enddata
\tablecomments{The format of EP Name for each high-confidence X-ray transient follows EPYYMMDDx, where x, starting with ‘a’, indicates the order in which the event was reported on that date. $T_0$ denotes the trigger time for confirmed X-ray transients. $S_X$ represents the WXT detection significance~\citep{yuan2022einstein}.}
\end{deluxetable*}

After applying the cross-matching procedure described in Section~\ref{sec:method}, we identified 16 candidates that satisfy our selection criteria, including 9 high-confidence eFXTs candidates and 7 \textit{unverified sources}, as listed in Tables~\ref{table:candidates_highconf} and~\ref{table:candidates_unverified}. 
Several candidates in our list have been extensively followed up and analyzed in the literature~\citep{liu2025soft,van2025einstein,sun2025fast,zhu2025ep250108a,hamidani2025ep240414a,li2025extremely,eyles2025ep250108ahop,srinivasaragavan2025ep250108a,wu2024ep241030a,jiang2025ep250226a}. For the remaining events, we carried out case-by-case inspections to evaluate the likelihood of a genuine association. Candidates lacking supporting multiwavelength evidence---such as UV or radio detections---or those coincident with catalogued X-ray sources possessing sub-$10^{\prime\prime}$ localizations are attributed to chance coincidence. For other events, we examined them individually to assess the likelihood of a genuine association. Those lacking multiwavelength evidence, such as UV/radio emission or catalogued X-ray sources with sub-$10^{\prime\prime}$ localization, will be attributed to chance coincidence. 
%Each candidate was examined individually to assess the likelihood of a genuine association and to exclude cases of chance spatial and temporal coincidence. 

% =====================
% Unverified sources
% =====================
\begin{deluxetable*}{ccccccc}[htb]
\small
\tablecaption{List of \textit{unverified sources} and identified optical counterpart candidates.}
\label{table:candidates_unverified}
\tablehead{
\colhead{EP Name} & \colhead{$T_{0}$} & \colhead{$S_{X}$} & \colhead{TNS Name} & \colhead{Classification} & \colhead{Separation} & \colhead{$\delta t$} \\ 
& & & & & [arcminutes] & [days]
}
\startdata
EPW20240326aa & 2024-03-24 08:00:27 & 5.27 & AT~2024fkr & Unverified & 0.72 & 9.052 \\ 
WXT~J1212+1004 & 2024-03-28 19:31:39 & 5.63 & AT~2024fyr & Unverified & 3.13 & 5.502 \\ 
WXT~J113313$-$003848 & 2025-04-18 03:18:39 & 5.05 & AT~2025kbw & Unverified & 3.02 & 5.514 \\ 
WXT~J071516+592950 & 2025-02-03 17:45:39 & 5.09 & AT~2025ccx & Unverified & 2.29 & 16.261 \\ 
WXT~J154218$-$072742 & 2025-04-30 23:46:13 & 5.06 & AT~2025lph & Unverified & 1.48 & 19.357 \\ 
WXT~J1454+1837 & 2024-05-02 16:02:29 & 5.28 & AT~2024ode & Unverified & 1.49 & 6.864 \\ 
WXT~J160346+193555 & 2024-06-11 01:28:15 & 5.59 & AT~2024loi & Unverified & 0.29 & 2.349 \\ 
\enddata
\tablecomments{$T_0$ denotes the observation start time for unverified sources.}
\end{deluxetable*}

We conclude that none of the \textit{unverified sources} are of extragalactic origin: WXT~J160346+193555 is likely associated with a CV, while the remaining six candidates are consistent with temporal or spatial coincidences (see Appendix~\ref{appendix:candidates}), and therefore fall outside the scope of this study. In addition to the well-studied eFXTs, we highlight EP240506a, which appears to be an \textit{orphan} event: despite extensive multiwavelength follow-up during the first few days, only upper limits were obtained. Remarkably, it is spatially and temporally associated with the optical transient AT~2024ofs, which was reported on TNS approximately 19 days after the EP trigger. We also identify prior UV emission at $\sim2$ days and early optical emission at $\sim8$ days post-trigger. Given this temporal offset, the chance-coincidence probability estimated in Section~\ref{sec:discuss} is $P_{\mathrm{cc}} \sim 3 \times 10^{-4}$, strongly suggesting that the association is genuine rather than a random alignment. 
%Interestingly, it is spatially and temporally associated with the optical transient AT~2024ofs, which was reported on TNS approximately 19 days after the EP trigger, and we found earlier UV emission at 2 days and early optical emission about 8 days after the WXT trigger. Given this temporal delay, the probability of chance coincidence derived in Section~\ref{sec:discuss} is $P_{\mathrm{cc}}\sim 3\times 10^{-4}$, indicating they are unlikely to be a random match. 
The WXT image and the position of AT\,2024ofs are shown in Figure~\ref{fig:pos}. Following a comprehensive analysis of archival data and the properties of its host galaxy, as detailed below, we confirm that EP240506a is associated with the SN candidate AT~2024ofs. This makes EP240506a another EP-detected eFXT linked to an SN.

\subsection{Observations and Analysis of EP240506a/AT\,2024ofs}
\label{subsec:Ep240506a}

In this section, we present the observations and analysis of the newly discovered SN candidate EP240506a/AT~2024ofs.

\subsubsection{X-ray/$\gamma$-ray Detection}\label{subsubsec:X_ray}

\begin{figure}[htb]
    \centering
    \includegraphics[width=0.99\linewidth]{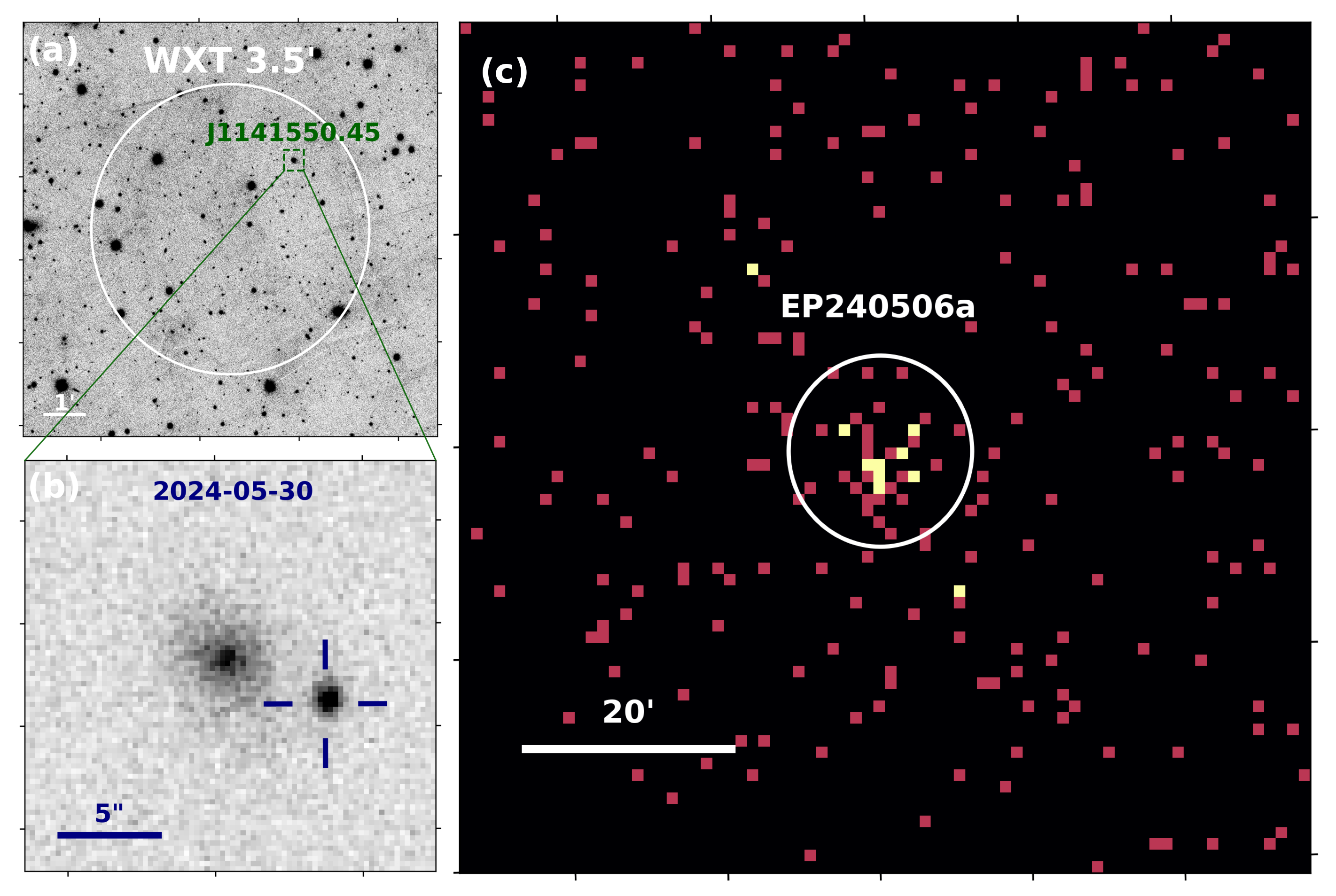}
    \caption{Localization of EP240506a and its optical counterpart AT\,2024ofs. (a) The position of the host galaxy J141550.45 (green outline) relative to the WXT localization uncertainty (white circle) in the PS $r$-band image. (b) The PS $w$-band science image, where the position of AT\,2024ofs is marked by a cross, showing that the transient lies on the outskirts of its host galaxy. (c) The WXT X-ray image obtained from $T_0$ to $T_0 + 150$~s. The white circle denotes the source extraction region with a radius of $9'$.}
    \label{fig:pos}
\end{figure}

EP240506a was discovered by WXT in the 0.5–4 keV band at 05:01:39 UTC on May 6, 2024, during the commissioning phase of EP~\citep{2024GCN.36405....1L}. The WXT data were processed with \texttt{wxtpipeline}, an analysis chain of the WXT Data Analysis Software (\texttt{WXTDAS}) developed by the EP Science Center (EPSC). The transient was detected with a significance of 7.8, which was below the on-board trigger\footnote{\textit{BeiDou} is the Chinese satellite navigation system. Alert data—including source coordinates, flux, spectral hardness ratio, and in some cases a simple light curve—are downlinked to the ground segment with a latency of $\sim$10~minutes. In addition, the CNES (France) VHF network, developed for the Chinese--French \textit{SVOM} mission and operating with a lower trigger threshold of 5.0, is used through collaboration to enhance the alert capability and enable the transmission of more extensive quick-look transient data~\citep{yuan2022einstein}.} threshold of 8.0 at that time, and therefore no automatic follow-up observations were triggered; the latter are essential for capturing the rapid decay phase.  Nevertheless, it was reported via the VHF Alert~\citep{yuan2022einstein}, with a sky position of R.A. = $14^{\mathrm{h}}\,15^{\mathrm{m}}\,56.88^{\mathrm{s}}$, Dec = $-16^{\circ}\,41^{\prime}\,16.8^{\prime\prime}$ (J2000), with position uncetainty of $3.5^{\prime}$. 

\begin{figure}[htb]
    \centering
    \includegraphics[width=0.8\linewidth]{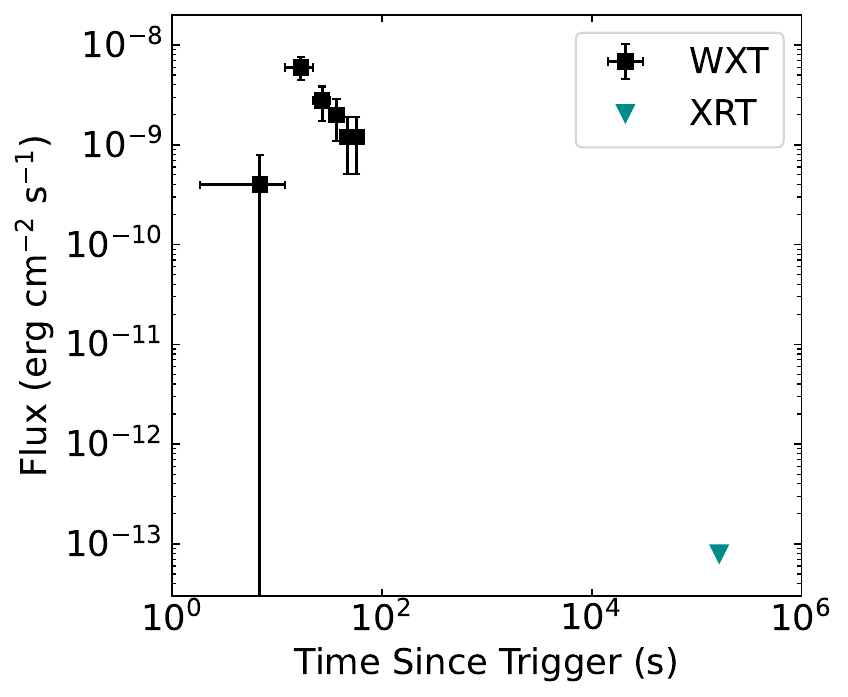}
    \caption{The 0.5-4.0\,keV light curve of EP240506a in the observer frame. The XRT flux is converted to 0.5-4~keV and is shown as a cyan triangle. The WXT data are binned with a time resolution of $10\,\mathrm{s}$. }
    \label{fig:X_lc}
\end{figure}

\begin{figure}[htb]
    \centering
    \includegraphics[width=0.9\linewidth]{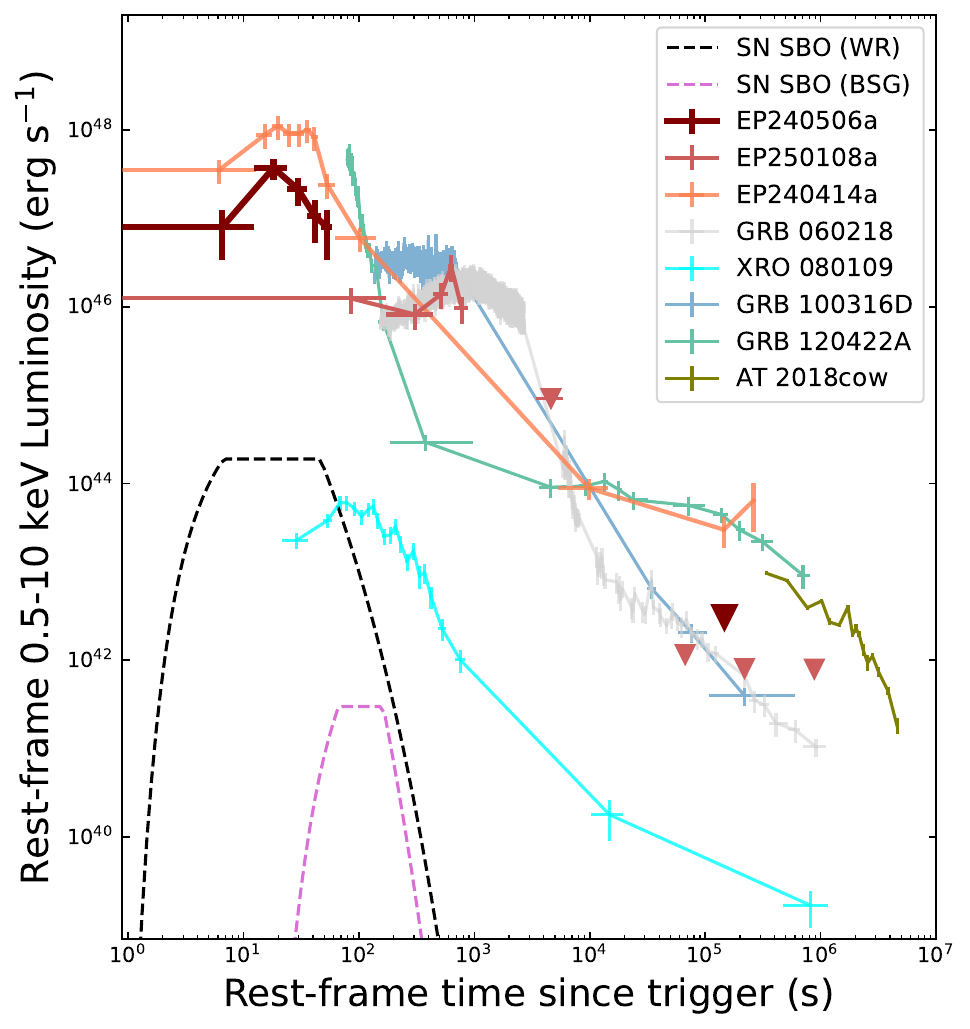}
    \caption{
    Rest-frame X-ray luminosity evolution of EP240506a, compared with those of EP240414a \citep{sun2025fast} and EP250108a \citep{li2025extremely}. The WXT count rates have been converted to 0.5--10~keV luminosities using the best-fit absorbed power-law model derived in Section~\ref{subsubsec:X_ray}, adopting $z=0.12$ (see Section~\ref{subsec:host}). Upper limits are shown as triangles. For comparison, we include the X-ray light curves of the low-luminosity GRBs 060218 \citep{campana2006association}, 100316D \citep{starling2011discovery}, and 120422A \citep{schulze2014grb}, obtained from the \textit{Swift} BAT and XRT data via the Burst Analyser \citep{evans2010swift}, with K-corrections applied using their time-resolved spectral fits. The dashed curves show the predicted shock-breakout luminosities for a Wolf–Rayet (WR) progenitor ($M=15\,M_{\odot}$, $R=5\,R_{\odot}$, $\kappa=0.2$) and a blue supergiant (BSG) progenitor ($M=15\,M_{\odot}$, $R=50\,R_{\odot}$, $\kappa=0.34$), following \citet{nakar2010early}. 
    % [Optional: Add description of a lower panel here if the figure has one, e.g., “The lower panel shows ...”]
    }
    %\caption{The rest-frame X-ray Luminosity of EP240506a, compared with EP240414a~\citep{sun2025fast} and EP250108a~\citep{li2025extremely}. The WXT data is converted to 0.5-10 keV assuming the best-fit powerlaw model obtained in Section~\ref{subsubsec:X_ray}. Triangle symbols indicate upper limits. The data of low-luminosity GRB 060218~\citep{campana2006association}, GRB 100316D~\citep{starling2011discovery}, and GRB 120422A~\citep{schulze2014grb} were obtained from BAT and XRT observations in the Swift Burst Analyser~\citep{evans2010swift}, with K-corrections applied using their time-resolved energy spectra. Dashed lines represent the X-ray luminosity of SBO from Wolf-Rayet (WR) with $M=15\ M_{\odot},\ R=5\ R_{\odot},\ \kappa=0.2$ and blue supergiant (BSG) with $M=15\ M_{\odot},\ R=50\ R_{\odot},\ \kappa=0.34$~\citep{nakar2010early}.\textcolor{red}{caption, parameters, lower panel}}
    \label{fig:L_lc}
\end{figure}

The flare, however, was missed in the default telemetry data as it occurred shortly after the passage through the South Atlantic Anomaly (SAA) and was truncated. The data were subsequently recovered by re-running the pipeline manually, yielding a duration exhibiting a duration of $T_{90} = 41.1^{+2.0}_{-4.0}\ \mathrm{s}$ and a peak flux of $\sim 1 \times 10^{-8}\ \mathrm{erg\ cm^{-2}\ s^{-1}}$ and total $36.8\pm 6.5$ photons, as the light curve shown in Figure~\ref{fig:X_lc}. The $T_{90}$ is defined as the time during which the central 90\% of the fluence is observed. The derived X-ray peak luminosity is $(2.3 \pm 0.6) \times 10^{47}\ \mathrm{erg\ s^{-1}}$ in the WXT 0.5--4.0\,keV band at $z=0.12$ (see Section~\ref{subsec:host}). 
The burst spectrum can be fitted by an absorbed power-law model with a photon index of $\Gamma = 0.85 \pm 0.44$, assuming a fixed Galactic hydrogen column density of $N_{\mathrm{H}} = 9.24 \times 10^{20}\ \mathrm{cm^{-2}}$~\citep{willingale2013nh}. The corresponding averaged unabsorbed flux is estimated to be $(1.61 \pm 0.41) \times 10^{-9}\ \mathrm{erg\ cm^{-2}\ s^{-1}}$ and corresponding averaged luminosity is $(6.45\pm 1.65)\times 10^{46}\ \mathrm{erg\ s^{-1}}$. As shown in Figure~\ref{fig:L_lc}, the X-ray luminosity of EP240506a significantly exceeds that of XRO~080109 and is comparable to GRB~060218, EP250108a, and EP240414a, implying the energy falls into the regime of relativistic SBO.

A \textit{Swift} follow-up observation commenced at 02:45:40 UTC on May 8, 2024, approximately 45 hours after the EP trigger, yielding a total exposure of 2.59\ ks. No uncatalogued X-ray source was detected within the $3.5'$ EP/WXT error circle. Assuming the best-fit absorbed power-law model derived for the prompt emission, we estimate a limiting flux of $7.9\times10^{-14}\ \mathrm{erg\ cm^{-2}\ s^{-1}}$ in the 0.5--4~keV band, calculated using the \texttt{HEASARC} \texttt{WebPIMMS} tool~\citep{mukai1993pimms}.
%No uncatalogued X-ray source was identified within the $3.5'$ error circle of EP/WXT. Assuming best-fit powerlaw model for prompt emission, the limiting flux of $7.9\times10^{-14}\ \mathrm{erg\cdot cm^{-2}\cdot s^{-1}}$ in the 0.5-4~keV band, which is derived using \texttt{HEASARC} tool \texttt{WebPIMMS}~\citep{mukai1993pimms}. 

Although the sky position of the transient was visible for \textit{Fermi}-GBM, no significant signal was detected within the time interval $T_{0} \pm 100\,\mathrm{s}$. By combining the non-detection from \textit{Fermi}-GBM down to $\sim 7\times 10^{-8}\ \mathrm{erg\ cm^{-2}\ s^{-1}}$ (assuming $\alpha=-0.85$, and $\beta=-2.3$) with the flux measured by WXT, and assuming the spectrum follows a \texttt{Band} function, we constrain the peak energy to $E_{\mathrm{peak}} < 140\ \mathrm{keV}$ and derive an isotropic-equivalent energy of $E_{\mathrm{iso}} < 9.3 \times 10^{49}\ \mathrm{erg}$, again adopting $z=0.12$ (see Section~\ref{subsec:host}). The derived upper limits on the peak energy and isotropic-equivalent energy fall within the region where Type-I and Type-II GRBs overlap on the Amati relation~\citep{amati2002intrinsic}. However, these limits rule out a large portion of the parameter space typically occupied by Type-I GRBs, providing meaningful constraints despite the remaining ambiguity.
%The derived upper limits on the peak energy and isotropic-equivalent energy lie within the parameter space of Type-II GRBs and do not strongly disfavor a Type-I GRB origin when evaluated against the Amati relation~\citep{amati2002intrinsic}.

%\begin{figure}
%    \centering
%    \includegraphics[width=0.8\linewidth]{Amati.jpg}
%    \caption{\textcolor{red}{need to modify} Amati relation}
%    \label{fig:amati}
%\end{figure}

\subsubsection{UV/Optical/NIR Observations}
\label{subsubsec:optical}

\begin{figure}[htb]
    \centering
    \includegraphics[width=1.\linewidth]{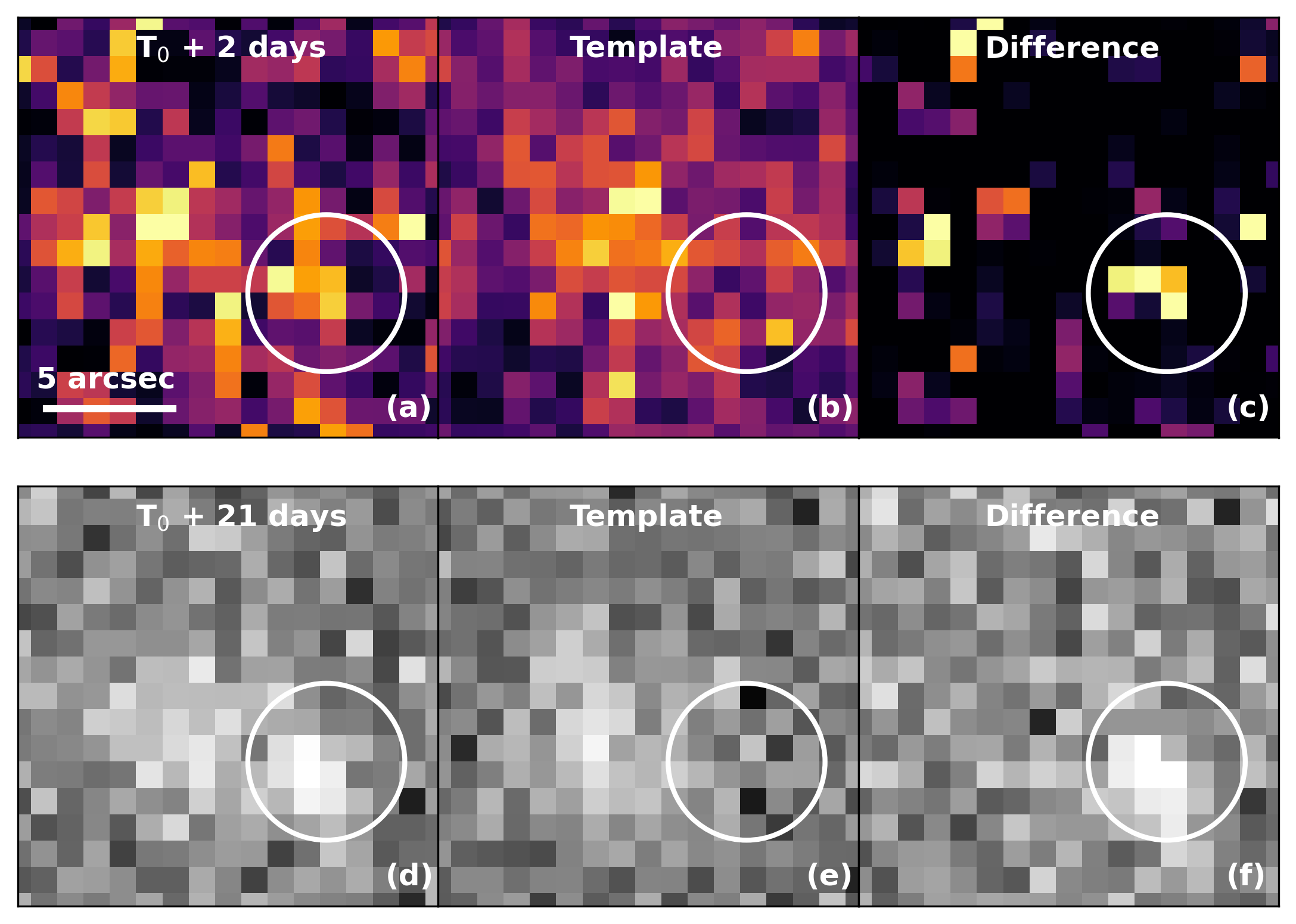}
    \caption{
    UVM2 and \textit{r}-band imaging of AT~2024ofs. 
    Panels (a)–(c) show the \textit{Swift}/UVOT UVM2 data: (a) the image obtained during the 2024 ToO observation, (b) the deeper template image taken in 2025, and (c) the difference image, displayed with a squared intensity scale. 
    Panels (d)–(f) show ZTF \textit{g}-band imaging: (d) the science image containing the transient, (e) the template image acquired on 8 May 2024, and (f) the corresponding difference image. 
    The position of AT~2024ofs is indicated by a white $3^{\prime\prime}$ radius circular aperture. All images are centered on the host galaxy. 
    }
    \label{fig:uvot_img}
\end{figure}

Analysis of the initial \textit{Swift}/UVOT ToO data show a marginal UVM2-band excess at the position of the optical counterpart, measured at $22.57 \pm 0.43$~mag (S/N = 2.5) in a 903.8\,s exposure (Figure~\ref{fig:uvot_img}a), with a $5\sigma$ limiting magnitude of $>21.59$~mag. This potential new detection, which has not been previously reported, lies near the sensitivity limit of the observation and may be affected by background fluctuations. To evaluate its reliability, we conducted a series of follow-up \textit{Swift} ToO observations between 2025 May 20 and 2025 June 24. A stacked UVM2-band image constructed with \texttt{uvotimsum} (total exposure of 2181.5\,s) achieved a $5\sigma$ limiting magnitude of $>22.31$ but revealed no significant emission at the position of AT\,2024ofs (Figure~\ref{fig:uvot_img}b). The absence of persistent emission in the deeper template suggests that the earlier excess, if real, was transient in nature; however, given its low significance, the possibility of a spurious fluctuation cannot be excluded. Nonetheless, we adopt this measurement in our analysis owing to its temporal and spatial consistency with the optical counterpart.

%Despite multiple follow-up campaigns following the EP trigger, no optical/IR counterpart was initially identified within the first 3 days.~\citep{2024GCN.36419....1J,2024GCN.36417....1P,2024GCN.36414....1C,2024GCN.36413....1T,2024GCN.36412....1P,2024GCN.36408....1A}. However, approximately two weeks later, a new optical transient, AT~2024ofs/ZTF24aaowcoo, was independently discovered by ZTF and Pan-STARRS within the \textcolor{red}{$3.5'$} localization uncertainty of the WXT detection~\citep{2024TNSTR2275....1C}. An early detection 7.7 days after WXT trigger emerged in the ATLAS forced photometry (see Figure~\ref{fig:optical_lv}). The transient is spatially coincident with the UVOT source and exhibits a fading trend as revealed by the ZTF Forced Photometry (see below)~\citep{masci2023new}.

Despite extensive follow-up campaigns within the first three days after the EP trigger, no optical/IR counterpart was identified~\citep{2024GCN.36419....1J,2024GCN.36417....1P,2024GCN.36414....1C,2024GCN.36413....1T,2024GCN.36412....1P,2024GCN.36408....1A,2025arXiv250421096A}. However, nearly two weeks later, a new optical transient, AT\,2024ofs/ZTF24aaowcoo, was independently discovered by ZTF and PS within the $3.5'$ localization uncertainty of the WXT detection~\citep{2024TNSTR2275....1C}. The transient is spatially coincident with the UVOT source and shows a fading trend in the ZTF forced-photometry light curve \citep[see below;][]{masci2023new}.  

From Figure~\ref{fig:pos}(b), AT\,2024ofs is offset by $5.32^{\prime\prime}$ from the galaxy SDSSJ141550.45$-$163937.0 (hereafter J141550.45), consistent with a location in the galaxy’s halo region. To further assess its origin, we revisited archival observations from multiple facilities—including ZTF, ATLAS, PS, GROND, Thai Robotic Telescope (TRT), and Small \& Light Telescope (SLT), and the Xinglong 2.16-m telescope (XLT)—and compiled them to reconstruct the optical light-curve evolution of AT\,2024ofs.  An even earlier signal was revealed in ATLAS forced photometry at 7.7 days after the WXT trigger. 

\begin{figure}[htb]
    \centering
    \includegraphics[width=0.97\linewidth]{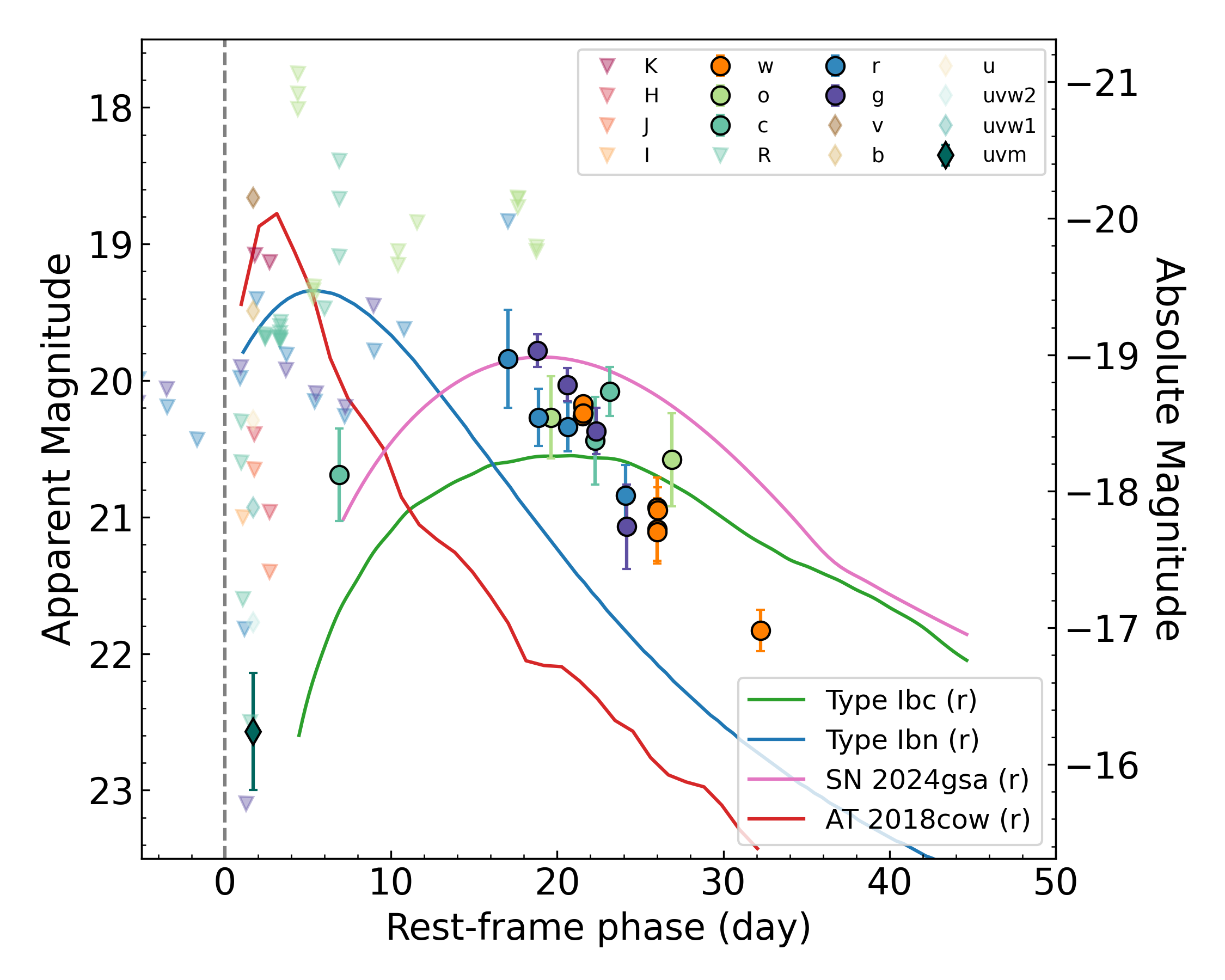}
    \caption{Optical light curve of AT\,2024ofs compared with $r$-band template absolute-magnitude light curves at $z=0.12$: SN~Ibc (green;~\citealt{2005ApJ...624..880L,drout2011first}), SN~Ibn (blue;~\citealt{hosseinzadeh2017type}), FBOT AT~2018cow (red;~\citealt{perley2019fast}), and Ic-BL SN~2024gsa (violet;~\citealt{sun2025fast}). The grey dashed line represents the WXT trigger time of EP240506a.}
    \label{fig:optical_lv}
\end{figure}

To improve the photometric quality, we employed the Forced Photometry Service~\citep{masci2023new} for ZTF data between $T_0 + 13$ days (MJD 60450.27), when the transient initially emerged on the subtracted ZTF image, and $T_0 + 28$ days (MJD 60464.27), which significantly enhanced the accuracy of the light curves. For the ATLAS light curves, we used the publicly available ATClean\footnote{https://github.com/srest2021/atclean} package \citep{Rest23atclean,Rest25atclean} to retrieve the forced photometry from the  ATLAS forced photometry\footnote{https://fallingstar-data.com/forcedphot/}  server \citep{Shingles21atlas} and clean it following the recipe described in \cite{Rest25atclean}. For PS data, photometric monitoring started on MJD 60460.327, 24.1 days after the discovery of EP240506a. The images were obtained in the \textit{w} filter \citep{2012ApJ...750...99T}. Images were processed with the Image Processing Pipeline \citep{2020ApJS..251....3M,Waters2020}, astrometrically and photometrically calibrated \citep{2020ApJS..251....6M}, and individual frames were coadded with median clipping to produce stacks, on which PSF photometry was performed \citep{2020ApJS..251....5M}. In addition, we co-added PS images taken around $T_{0}+35.8$ days (MJD 60472), yielding a clear detection of the transient at $w = 21.83 \pm 0.15$ mag using \textsc{AutoPhOT}. The source subsequently faded and was not detected in later epochs. 

The sky position of AT\,2024ofs was also observed as part of our $J$-band follow-up of EP\,240506a using the 7-channel imager GROND \citep[][]{Greiner2008a} mounted on the MPG 2.2\,m telescope at ESO’s La Silla Observatory. The observation was carried out on 2024 May 9 at UT 05:02, corresponding to 72 hours after the EP discovery. Image differencing was performed using the \textsc{AutoPhOT} pipeline \citep{Brennan2022d}, with a $J$-band image from the VISTA Hemisphere Survey\footnote{\url{https://www.eso.org/sci/observing/phase3/data_releases/vhs_dr1.html}} obtained in May 2012 serving as the reference. No source was detected at the target position, with 5-$\sigma$ limiting Vega magnitudes of $J>19.1,\ H>18.3$, and $K>17.3$.  

The 40~cm SLT at the Lulin observatory was utilized to obtain images in the SDSS $r$ filter as part of the Kinder collaboration \citep[][]{2025ApJ...983...86C}. We co-added $30 \times 300$\,s images from SLT, reaching a $3\sigma$ limiting magnitude of $\sim 22$ in the $r$ band.

The collected light curve is presented in Figure~\ref{fig:optical_lv}, which we contrast against the light curves of SN~Ibn and SN~Ibc templates, as well as the prototypical FBOT AT\,2018cow and Ic-BL SN\,2024gsa associated with EP240414a. The evolution of AT\,2024ofs exhibits distinct features relative to these classes: it shows a longer rise time but a comparable post-peak decay rate to AT\,2018cow and SNe~Ibn, along with a lower peak luminosity. Compared to typical SNe~Ibc (as well as SN\,2024gsa), it evolves more rapidly, indicating that AT\,2024ofs belongs to a class of intermediate-timescale transients rather than representing the extremes of either rapidly evolving or normal stripped-envelope SNe.

\subsubsection{Multi-wavelength Fitting}

\begin{figure}[htb]
    \centering
    \includegraphics[width=0.97\linewidth]{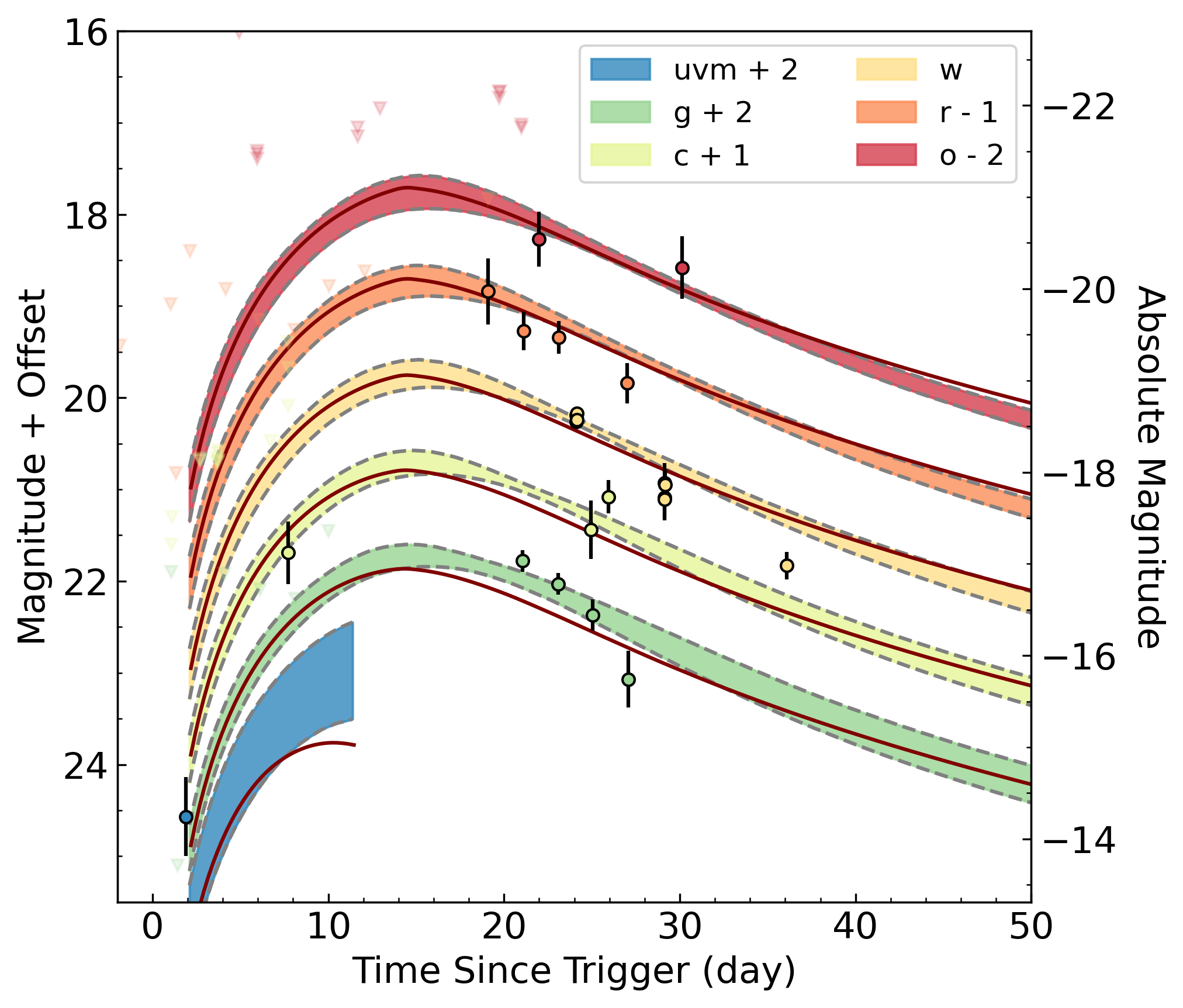}
    \caption{Multiband optical light curves of AT~2024ofs compared with the best-fit Arnett model. Observed photometric data with error bars are shown together with the 90\% credible intervals of the model predictions (shaded regions). The UVM2-band data and corresponding model prediction are also shown to illustrate the deviation at early times. The red curves represent a model with parameters $M_{\mathrm{Ni}} = 0.8\,M_{\odot}$, $M_{\mathrm{ej}} = 2.5\,M_{\odot}$, and $v_{\mathrm{ej}} = 1.6 \times 10^{4}\,\mathrm{km\,s^{-1}}$. Different filters are indicated by distinct colors and vertical offsets for clarity, where non-detections are plotted as triangles. The left axis denotes apparent magnitudes, while the right axis shows the corresponding absolute magnitudes.}
    \label{fig:lc_fit}
\end{figure}

The observed data were corrected for a line-of-sight Galactic extinction of $E(B-V)=0.106$\,mag~\citep{schlegel1998map}. The UV detection was excluded from the fit because it deviates significantly from the model light curve (see Figure~\ref{fig:lc_fit}) as well as its low $S/N$, and because the fitting framework may not be applicable at such early epochs.
%The UV detection was excluded from the fit because it deviates significantly from the predicted light curve (see Figure~\ref{fig:lc_fit}), likely due to an additional component such as shock cooling of the stellar envelope or an afterglow from a mildly relativistic cocoon or an off-axis jet~\citep{nakar2010early,waxman2017shock,forster2018delay}.}

We modeled the rest of the multi-wavelength light curve of AT~2024ofs using the semi-analytical Arnett model~\citep{arnett1982type} as implemented in the \texttt{Redback} package\footnote{\url{https://github.com/nikhil-sarin/redback}}~\citep{sarin2024redback}. The model assumes homologous ejecta expansion, with heating powered by the radioactive decay of ${}^{56}$Ni and ${}^{56}$Co, while photon diffusion regulates the emergent luminosity~\citep{arnett1982type}. Bayesian inference was performed via nested sampling with \texttt{dynesty}\footnote{\url{https://github.com/joshspeagle/dynesty}}~\citep{speagle2020dynesty}, using a likelihood function that incorporates both detections and non-detections. The analysis highlights the sparse temporal coverage—particularly the lack of photometry around peak luminosity—which produced poorly constrained posteriors. Although the model broadly reproduces the observed light curves (Figure~\ref{fig:lc_fit}), the inferred parameters drift toward unphysical regions of parameter space, as illustrated in Figure~\ref{fig:corner} in Appendix~\ref{appendix:fitting}.

For context, we compare our results with ejecta properties inferred for other SNe associated with eFXTs detected by EP~\citep{sun2025fast,li2025extremely}, which indicate ${}^{56}$Ni masses of $\sim0.8\ M_{\odot}$ and relatively high $M_{\mathrm{Ni}}/M_{\mathrm{ej}}$ ratios. Assuming this ${}^{56}$Ni mass, the expected peak bolometric absolute magnitude is $M_{\mathrm{peak}}=-(\mathrm{log_{10}}M_{\mathrm{Ni}}+0.8184)/0.415=-19.5$, which is consistent with our observations~\citep{2016MNRAS.457..328L}. Motivated by these studies, we fixed the ${}^{56}$Ni mass to a plausible value and allowed other parameters to vary within reasonable ranges, ensuring that the model predictions remained broadly consistent with the limited data. While this approach yields a qualitative description of the event, it should not be regarded as unique or statistically robust.

The optical light curves are reasonably reproduced by a radioactive-decay-powered model, yielding an ejecta mass of $M_{\mathrm{ej}}\approx2.5\ M_{\odot}$, $v_{\mathrm{ej}} = 1.6 \times 10^{4}\,\mathrm{km\,s^{-1}}$, and kinetic energy of $E_{\mathrm{K}}=3v_{\mathrm{sc}}^{2}M_{\mathrm{ej}}/10 \approx 1.9\times10^{51}\ \mathrm{erg}$. The photospheric velocity at peak light, $v_{\mathrm{sc}}$, and the peak bolometric luminosity, $L_{\mathrm{peak}}\approx2.3\times10^{43}\ \mathrm{erg\ s^{-1}}$, are inferred from the fitted parameters. These values are broadly consistent with those of similar events, although the adopted ${}^{56}$Ni mass is higher than that of typical stripped-envelope SNe. However, it remains compatible with the yields of energetic SN~Ic-BL given the high peak luminosity, supporting the interpretation that the emission was primarily powered by ${}^{56}$Ni and ${}^{56}$Co decay~\citep{campana2006association,soderberg2008erratum,sun2025fast,li2025ep250108a,2016MNRAS.457..328L,2016MNRAS.458.2973P,2019A&A...621A..71T}.
%These values are broadly consistent with previous similar events~\citep{campana2006association,soderberg2008erratum,sun2025fast,li2025ep250108a}, supporting the interpretation that the emission was primarily powered by ${}^{56}$Ni and ${}^{56}$Co decay.

Nonetheless, these estimates remain subject to large uncertainties because of the sparse temporal sampling and the absence of spectroscopy for AT~2024ofs. Future events with denser photometric coverage and prompt spectroscopy will be crucial for placing robust constraints on explosion mechanisms and energy sources.

\subsubsection{Host Galaxy}\label{subsec:host}

%To determine the redshift of EP240506a/AT~2024ofs, we obtained the spectrum of its host galaxy using FORS2, mounted on the Very Large Telescope (VLT) at the European Southern Observatory (ESO), considering the transient had already faded by the time of observation. We calibrated the flux by multiplying by a scale factor to be consistent with photometry data, because only a fraction of photons are collected in the Long Slit Spectrum (LSS) for an extended source. The host galaxy spectrum exhibits a continuum with prominent emission lines, including $\mathrm{H\alpha}$ and $\mathrm{[N\ II]}$ as shown in Figure~\ref{fig:host_spec}, enabling a reliable redshift measurement of $z=0.120 \pm 0.002$. \textcolor{red}{host extinction here}. 

\begin{figure}[ht!]
    \centering
    \includegraphics[width=0.99\linewidth]{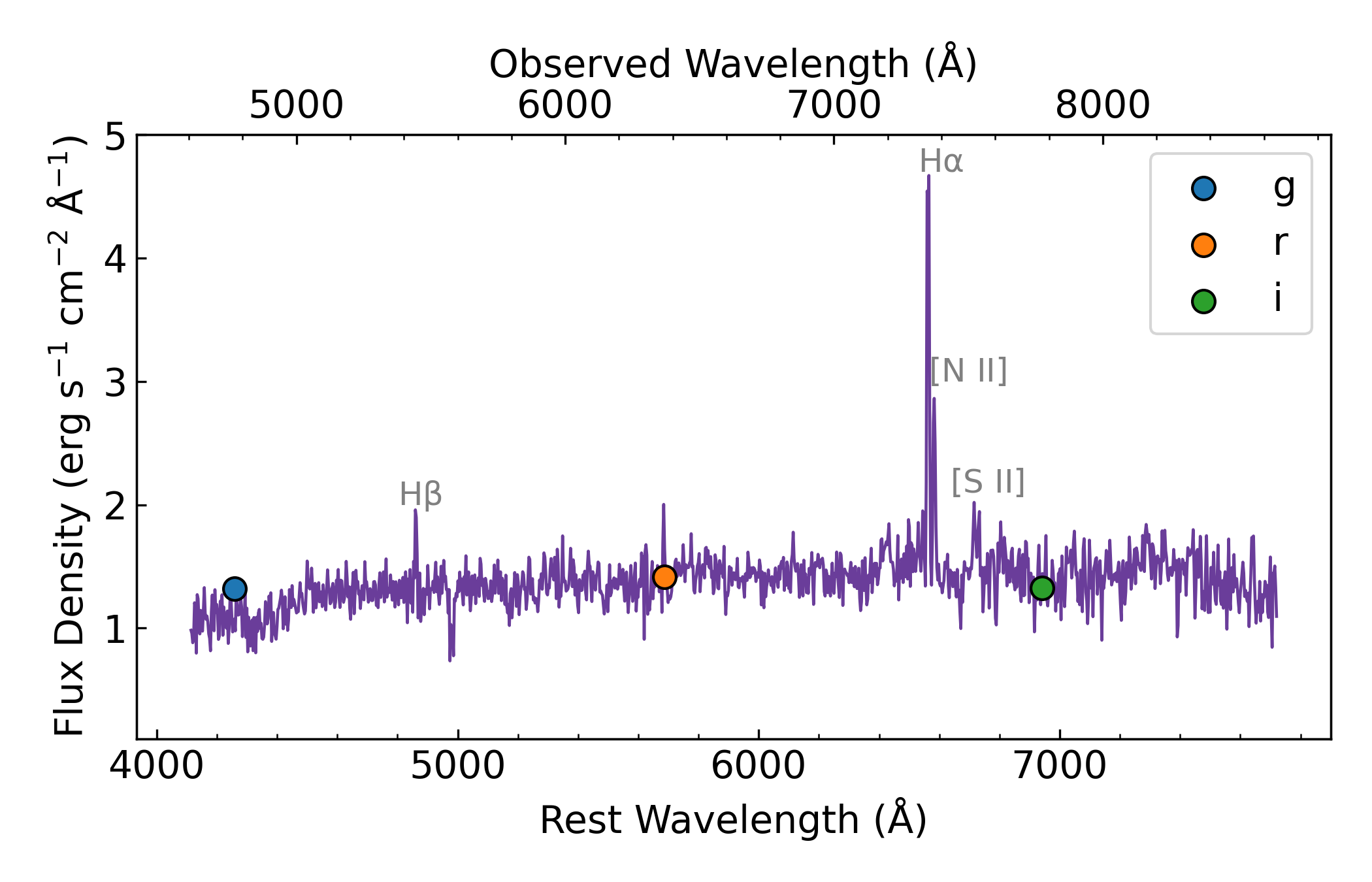}
    \caption{Optical spectrum of the host galaxy. The flux calibration was performed using photometric measurements from the Legacy Survey, shown as colored dots. Prominent emission lines employed for redshift measurement are indicated with gray labels.}
    \label{fig:host_spec}
\end{figure}

To determine the redshift of EP240506a/AT~2024ofs, we obtained a long-slit spectrum of its host galaxy with the FOcal Reducer/low dispersion Spectrograph 2 (FORS2) on the ESO VLT (UT1) at Paranal on 2025 April 26 (Program ID: 115.287Q.002, PI: Ling-Zhi~Wang). The data were processed with the ESO Reflex workflow \citep{2013A&A...559A..96F} using standard procedures (bias subtraction, flat-fielding, wavelength calibration, and sky subtraction).%given that the transient had already faded by the time of observation. 
The flux calibration was performed by applying a scaling factor to match the photometric measurements, as the spectroscopic extraction encompassed only a fraction of the flux from this extended source. The host galaxy spectrum exhibits a continuum with prominent emission features, including $\mathrm{H\alpha}$ and $\mathrm{[N\ II]}$, as shown in Figure~\ref{fig:host_spec}.  By fitting the emission lines, we derived a redshift of $z = 0.120 \pm 0.002$. We then performed full spectral fitting using the method of \citet{Li_N_2020}. Prior to fitting, all detected emission lines were carefully masked following the procedure outlined in \citet{Li_C_2005}. Our fitting employs the simple stellar population (SSP) models of \citet{Bruzual_Charlot_2003}, which comprise 1326 SSPs spanning 221 ages from 0 to 20 Gyr and six metallicities from $Z = 0.005 Z_\sun$ to $Z = 2.5 Z_\sun$. These models are based on the Padova evolutionary tracks \citep{Bertelli_G_1994} and a \citet{Chabrier_G_2003} stellar initial mass function. From this fitting, we derive a dust attenuation of the host galaxy of $E(B-V)=0.19$.

Assuming the cosmological parameters from the \textit{Planck} Collaboration \citep{aghanim2020planck}, the luminosity distance and angular-diameter distance to the host galaxy are 578.5 and 461.2~Mpc, respectively. Given an angular offset of $5.32^{\prime\prime}$, the projected physical separation between EP240506a/AT2024ofs and the center of its host galaxy is estimated to be 11.9~kpc.

We also estimated the probability of a chance coincidence between AT\,2024ofs and J141550.45. Using the apparent r-band magnitude of 18.2 mag from the Legacy Survey and the angular separation of $5.32^{\prime\prime}$, we derived a chance-coincidence probability of $P_{\mathrm{ch}} = 4.8 \times 10^{-3}$ following the method of \citet{bloom2002observed}. Such a low probability suggests that the association between the SN and the galaxy is unlikely to be a random alignment.

\subsection{Probability of Chance Coincidence}
\label{subsec:pcc}

%Our search for optical counterparts to eFXT candidates detected by EP yielded a multitude of matches. However, only a few are unlikely to be chance associations given the large positional uncertainties of WXT, which represent the dominant source of contamination in our search. It is therefore crucial to establish a robust assessment of the probability of random coincidence in order to estimate the expected number of true associations as the sample continues to grow. 
Our search for optical counterparts to eFXT candidates yields more than ten matches. However, the relatively large localization uncertainties of WXT imply that a substantial fraction of these associations are consistent with chance alignments, which dominate the contamination. Quantifying the rate of chance coincidences (i.e., false associations) is therefore essential for estimating the expected number of genuine counterparts as the sample size increases.

\begin{figure}[htb!]
    \centering
    \includegraphics[width=0.8\linewidth]{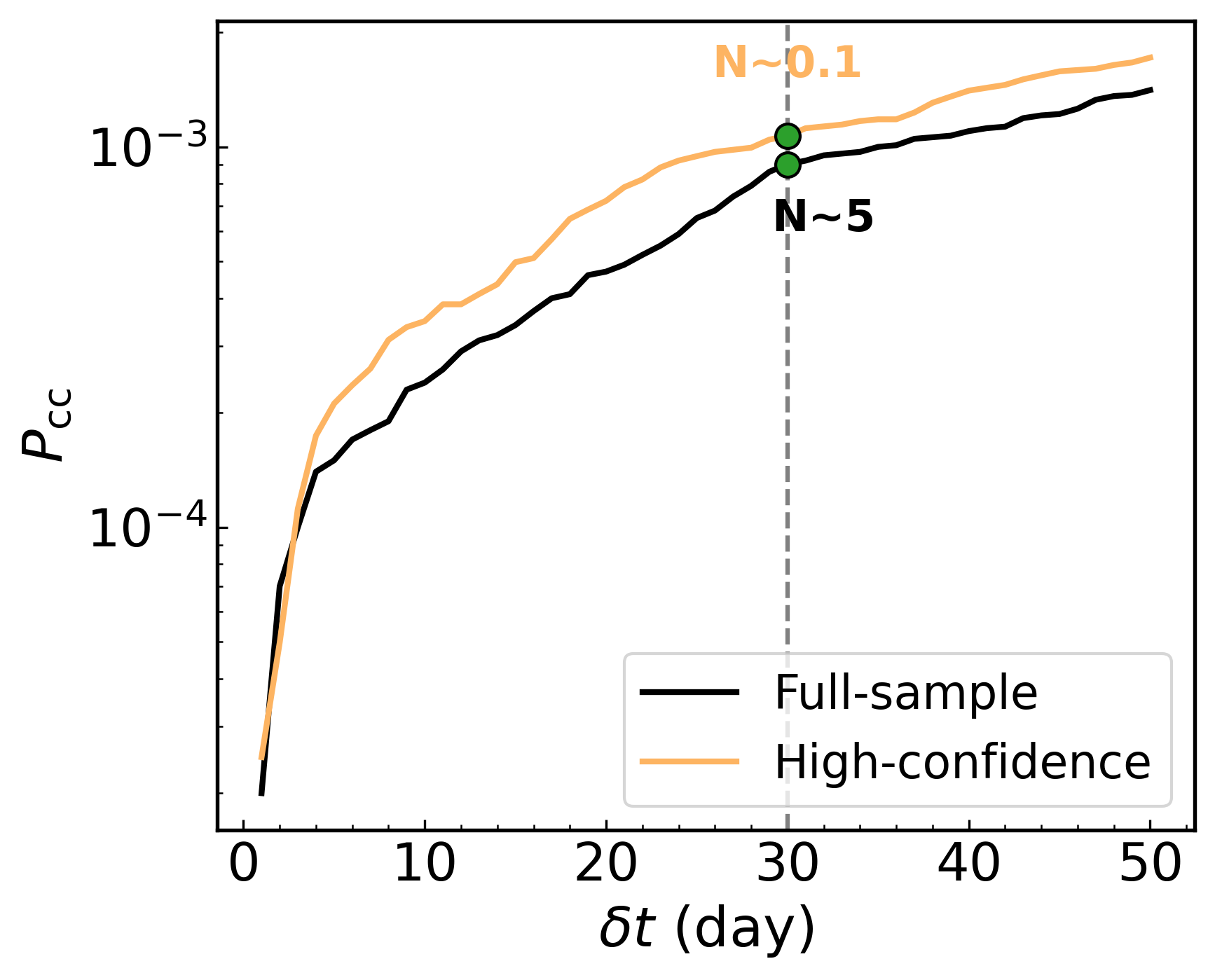}
    \caption{$P_{\mathrm{cc}}$ versus time offset $\delta t$ for the full EP eFXT candidate sample (black) and the high-confidence eFXT candidate subsample (orange). The vertical dashed line marks $\delta t=30$\,days, the threshold adopted in this work. The green dots denote the expected random matches at $\delta t=30$\,days.}
    \label{fig:pcc}
\end{figure}

To evaluate the likelihood of X-ray--optical match arising from chance coincidence ($P_{\mathrm{cc}}$), we construct a mock sample of optical transients by resampling the discovery dates from the real optical sample in Section~\ref{sec:method} and applying a random positional offset within an annulus with inner and outer radius of $3.5^{\prime}$ and $1^{\circ}$, to prevent true association in our simulation. This approach introduces spatial and temporal randomness while preserving the underlying sky and time distributions. We then performed one-dimensional Kolmogorov–Smirnov (KS) tests on the distributions of discovery date and sky position ($t_{\mathrm{diso}}$, RA, Dec) for the real and mock samples. The null hypothesis that the two samples are drawn from the same distribution could not be rejected, with $p$-values well above 0.05 (corresponding to a 90\% confidence level).

%We estimate $P_{\mathrm{cc}}$ via 200 Monte Carlo simulations \citep{2011A&A...530A..72M,2021arXiv210801805V,2022ApJ...939L..14W}. In each simulation, we randomly select 500 X-ray sources ($\approx 10\%$ of the full EP~eFXT candidate catalog) and cross-match them with the mock optical sample using the procedure described in Section~\ref{sec:method}. $P_{\mathrm{cc}}$ is defined as the fraction of X-ray sources with at least one counterpart in the mock optical sample. By varying the temporal offset thresholds $\delta t$, we derive the $P_{\mathrm{cc}}$ as a function of $\delta t$ for both the full eFXT candidate sample and the subset of high-confidence candidates, as shown in Figure~\ref{fig:pcc}. The lower $P_{\mathrm{cc}}$ measured for the full eFXT candidate sample is driven by the inclusion of Galactic-plane sources: the mock optical transients are concentrated at high Galactic latitude, reducing the sky overlap and thus the chance-coincidence rate, as seen in Figure~\ref{fig:sky}. At $\delta t=30$\,days, the expected number of random matches is $\sim10$, consistent with the seven unrelated sources identified in this study (see Section~\ref{subsec:candidates} and Appendix~\ref{appendix:candidates}). And as seen, the time delays for high-confidence eFXTs, which are always within ten days, are quite smaller than sub-threshold sources thanks to rapid multi-wavelength follow up observations. Therefore, we estimate the expected random match for high-confidence eFXTs is around 0.2 with $\delta=10$~days. 

We estimate $P_{\mathrm{cc}}$ via 200 Monte Carlo simulations \citep{2011A&A...530A..72M,2021arXiv210801805V,2022ApJ...939L..14W}. In each simulation, we randomly select 500 X-ray sources ($\approx 10\%$ of the full EP eFXT candidate catalog) and cross-match them with the mock optical sample following the procedure described in Section~\ref{sec:method}. We define $P_{\mathrm{cc}}$ as the fraction of X-ray sources that yield at least one counterpart in the mock optical sample. By varying the temporal offset threshold $\delta t$, we derive $P_{\mathrm{cc}}(\delta t)$ for both the full eFXT candidate sample and the subset of high-confidence events, as shown in Figure~\ref{fig:pcc}.

The lower $P_{\mathrm{cc}}$ inferred for the full eFXT candidate sample is primarily driven by the inclusion of X-ray sources located near the Galactic plane: because the mock optical transients are distributed predominantly at high Galactic latitude, the effective sky overlap is reduced, leading to a lower chance-coincidence rate (see Figure~\ref{fig:sky}). Adopting $\delta t = 30$\,days, the expected number of random matches is $\sim 5$, fully consistent with the seven unrelated associations identified in this work (see Section~\ref{subsec:candidates} and Appendix~\ref{appendix:candidates}). 

For the subset of high-confidence eFXTs, we estimate an expected number of random associations of only $\sim 0.1$ at $\delta t = 30$\,days. These events generally exhibit much shorter X-ray–optical time delays—typically within ten days—owing to rapid multi-wavelength follow-up efforts. Such prompt observations substantially suppress the probability of spurious positional and temporal coincidences.

\section{Discussion}\label{sec:discuss}

Here, we discuss the implications of the discovery of the optical counterpart of EP240506a and the estimation of local event rate for EP240506a-like events. 

The discovery of EP240506a/AT~2024ofs illustrates how SN-associated eFXTs can be missed in real time, even when the association is clear in hindsight. The event significance (\(S/N = 7.8\)) fell just below the on-board trigger threshold, preventing an automated FXT follow-up and the acquisition of a precise X-ray localization within the first day. Although dedicated follow-up observations were conducted within the first \(\sim 3\) days, no optical counterpart was detected, likely owing to the delayed and relatively slow optical rise of the SN. Without an early detection to motivate continued monitoring, the optical rise and peak were entirely missed. This case highlights how coarse localizations and short-lived follow-up can obscure SN counterparts to eFXTs, even for nearby and intrinsically luminous events.

More broadly, EP240506a/AT~2024ofs supports the emerging picture that a population of low-luminosity or cocoon-powered collapsars contributes to the EP eFXT sample, tracing a continuum between classical long GRBs and ordinary core-collapse SNe. This discovery underscores the importance of rapid X-ray localization and sustained optical monitoring for fully characterizing this population.

%\subsection{Rates}

To place EP240506a-like events in a broader population context, we estimate the local event rate density of eFXT analogous to EP240506a following~\cite{sun2015extragalactic}. The local event rate density can be expressed as

\begin{equation}
    \rho_{0} = \frac{4\pi N}{\eta \Omega_{\mathrm{WXT}}T_{\mathrm{OT}}} \frac{1}{V_{\mathrm{max}}},
\end{equation}

\noindent where $N$ is the number of detected events, $\Omega_{\mathrm{WXT}}=3600~\mathrm{deg}^{2}$ is the field of view of EP/WXT, $\eta\simeq 0.5$ is the duty cycle~\citep{sun2025fast}, and $T_{\mathrm{OT}}\simeq1.5~\mathrm{yr}$ is the time span covered by our search. The $V_{\mathrm{max}}$ represents the maximum volume within which the event can be detected, 

\begin{equation}
    V_{\max}=\int_{0}^{z_{\max}}\frac{dV_{c}}{dz}\,\frac{f(z)}{1+z}\,dz. 
\end{equation} 

\noindent The co-moving volume is given by 

\begin{equation}
    \frac{dV_{c}}{dz}
    =
    \frac{c}{H_{0}}\,
    \frac{4\pi\,D_{\mathrm{L}}(z)^{2}}{(1+z)^{2}\,E(z)}\,,
\end{equation}

\noindent with $E(z)=[\Omega_{\mathrm{M}}(1+z)^{3}+\Omega_{\Lambda}]^{1/2}$ for a flat $\Lambda$CDM cosmology \citep{aghanim2020planck}. Here $f(z)$ is the redshift-dependent evolution of star-formation history~\citep{yuksel2008revealing} and $D_{\mathrm{L}}$ is the luminosity distance at the corresponding redshift. 
Based on redshifted light-curve simulations, an EP240506a–like event with a peak X-ray luminosity of \((2.3\pm0.6)\times10^{47}\ \mathrm{erg\,s^{-1}}\) would be detectable by WXT out to \(z_{\max}=0.16\) at an S/N of 7. This implies a local event rate density of \(\rho_{0}=8.8^{+21.2}_{-3.9}\ \mathrm{yr^{-1}\,Gpc^{-3}}\), where the uncertainties represent the \(1\sigma\) Poisson limits \citep{gehrels1986confidence}. 

%Accounting for the fact that roughly one third of high-confidence eFXTs currently have reliable redshift measurements (Wu et al. in prep), we obtain a completeness-corrected rate of \(26.4^{+63.6}_{-11.7}\ \mathrm{yr^{-1}\,Gpc^{-3}}\). 
%This inferred rate is orders of magnitude below the core-collapse supernova rate of \((6.88 \pm 0.078) \times 10^{4}\ \mathrm{yr^{-1}\,Gpc^{-3}}\) \citep{2025A&A...698A.306M}, and is broadly consistent with the rates estimated for EP250108a \citep{li2025extremely} and for the population of low-luminosity GRBs (\(164^{+98}_{-65}\ \mathrm{yr^{-1}\,Gpc^{-3}}\); \citealt{sun2015extragalactic}).

As of June 2025, four EP transients have been associated with SNe—EP240414a, EP240506a, EP250108a, and EP250304a~\citep{sun2025fast,li2025extremely,Chen250304a,zhang2025ep250304a}. By combining these events \citep{Sun2022}, we estimate the local event rate density of EP-discovered SNe of about %$\rho_{0} = 14.8^{+11.9}_{-2.9}\ \mathrm{yr^{-1}\ Gpc^{-3}}$
$12$--$26\ \mathrm{yr^{-1}\ Gpc^{-3}}$. Accounting for the fact that roughly one third of high-confidence eFXTs currently have reliable redshift measurements (Wu et al. in prep), we obtain a completeness-corrected rate of %$44.3^{+35.8}_{-8.6}\ \mathrm{yr^{-1}\ Gpc^{-3}}$
$36$--$78\ \mathrm{yr^{-1}\ Gpc^{-3}}$. This inferred rate is orders of magnitude below the core-collapse supernova rate of \((6.88 \pm 0.078) \times 10^{4}\ \mathrm{yr^{-1}\,Gpc^{-3}}\) \citep{2025A&A...698A.306M}, and is broadly consistent with the rates estimated for EP250108a \citep{li2025extremely} and for the population of low-luminosity GRBs (\(164^{+98}_{-65}\ \mathrm{yr^{-1}\,Gpc^{-3}}\); \citealt{sun2015extragalactic}). 
%\textcolor{red}{While these inferred rates are broadly consistent with the rates estimated for EP250108a \citep{li2025extremely} and for the population of low-luminosity GRBs (\(164^{+98}_{-65}\ \mathrm{yr^{-1}\,Gpc^{-3}}\); \citealt{sun2015extragalactic}), they are orders of magnitude below the CCSNe rate of \((6.88 \pm 0.78) \times 10^{4}\ \mathrm{yr^{-1}\,Gpc^{-3}}\) \citep{2025A&A...698A.306M}, of which $\sim4\%$ are SN Ic-BL~\citep{Shivvers2017rate}, implying only a small fraction of SN Ic-BL could produce high-energy emission~\citep{corse2023search}.}
%and only $<19\%$ could launch low-luminosity GRBs

\section{Conclusion}\label{sec:conclusion} 

In this work, we conduct a systematic all-sky search for optical counterparts to a selected X-ray sample of EP, consisting of high-confidence eFXT candidates and \textit{unverified sources} identified by EP/WXT. The search adopts the following matching criteria: (1) spatial separation $\leq 3.5^{\prime}$ and (2) temporal offset between the X-ray and optical detections $\delta t \leq 30$~days.

Applying these criteria, we identify optical counterpart candidates for nine high-confidence eFXT candidates and seven \textit{unverified sources}. 
The majority of the high-confidence eFXTs are associated with either GRBs or supernovae, which have been well studied. 
Within this sample, we uncover one previously unrecognized EP–SN association, EP240506a.
%Among the \textit{unverified sources}, all but WXT~J160346+193555 (Appendix~\ref{J160346})—which is likely associated with a CV—are consistent with chance coincidences. 

AT\,2024ofs is a newly identified SN candidate associated with the eFXT event EP240506a, which has a duration of $T_{90} = 41.1^{+2.0}_{-4.0}\ \mathrm{s}$ and a peak X-ray luminosity of $(2.3 \pm 0.6) \times 10^{47}\ \mathrm{erg\ s^{-1}}$. No UV/optical counterpart had been previously reported for this event.  The \textit{Swift}/UVOT ToO data taken two days after the eFXT event revealed a low-significance detection (Figure~\ref{fig:uvot_img}(c)). AT~2024ofs was identified in optical imaging 7.8 days after the X-ray trigger at the same position as the UV source. Given the very low chance-alignment likelihood, we adopt AT~2024ofs as the optical counterpart to EP240506a. 

Using VLT/FORS2, we measured a spectroscopic redshift of $z=0.12$ for the host galaxy from prominent H$\alpha$ and [O\,\textsc{iii}] emission lines. %The host-galaxy extinction was estimated to be $E(B-V)=0.19$ from SED fitting with SSP models. 
The corresponding luminosity and angular-diameter distances are $D_L=578.5$ and $D_A=461.2$\,Mpc, respectively. 

We compiled archival multiwavelength photometric data for EP240506a/AT~2024ofs from initial follow-up observations and surveys, including ZTF, ATLAS, PS, GROND, XLT, TRT, and SLT. %We then applied forced photometry for coadded marginal detections to enhance the S/N and depth. 
The combined light curve can be broadly reproduced by radioactive decay model with ${}^{56}\mathrm{Ni}$ mass $\sim 0.8\ M_{\odot}$, ejecta mass $M_{\mathrm{ej}}=2.5\ M_{\odot}$, $v_{\mathrm{ej}} = 1.6 \times 10^{4}\,\mathrm{km\,s^{-1}}$, consistent with those of EP240414a/SN\,2024gsa and EP250108a/SN\,2025kg. Nevertheless, the uncertainties remain substantial, underscoring the need for future events with denser photometric sampling and timely spectroscopic observations to robustly constrain their physical origins.

We estimate the local volumetric event rate density for EP240506a-like transients. By simulating its light curve, we derive a maximum detectable redshift of $z_{\mathrm{max}}=0.16$ at an S/N of 7, corresponding to a event rate density of $\rho_{0}=8.8^{+21.2}_{-3.9}\ \mathrm{yr^{-1}\,Gpc^{-3}}$. If EP240506a arises from the core collapse of a massive star, combining this event with other EP-detected SN-associated transients yields a rough completeness-corrected population rate of $36$--$78\ \mathrm{yr^{-1}\ Gpc^{-3}}$.

%We further estimate $P_{\mathrm{cc}}$ in the cross-match process using Monte Carlo simulations. The expected numbers of random matches for the full X-ray sample and for the high-confidence eFXT candidates are $\sim5$ and $\sim0.1$, respectively. This suggests that most of the \textit{unverified sources} likely result from chance associations, consistent with our results (Appendix~\ref{appendix:candidates}). 

%Our results highlight the key role of systematic, multiwavelength counterpart searches for EP-discovered transients. Such multiwavelength campaigns are essential not only for revealing the nature of individual events but also for building a well-characterizsed population that will ultimately clarify the physical origins, diversity, and rates of eFXT transients. It is also important to systematically search for counterparts of eFXT detected by EP/FXT. Although the FoV of FXT ($\sim 1\ \mathrm{deg}^{2}$) is much smaller than that of WXT, its observable volume is $\sim 10\%$ that of WXT due to its significantly higher sensitivity. Moreover, the superior localization accuracy of FXT ($\sim 10^{\prime\prime}$) and its higher spectral resolution are critical for robust multiwavelength identification. Real-time counterpart searches are also essential for enabling prompt multiwavelength follow-up observations. 

Our results underscore the importance of systematic multiwavelength counterpart searches for EP-discovered transients. While EP/WXT provides wide-field discovery, searches for counterparts of EP/FXT detections are equally critical: despite its smaller field of view ($\sim1\ \mathrm{deg}^{2}$), FXT probes an observable volume of $\sim10\%$ that of WXT owing to its higher sensitivity and delivers much improved localization ($\sim10^{\prime\prime}$) and spectral quality. Real-time counterpart searches and prompt multiwavelength follow-up are crucial to fully exploit the scientific potential of EP.

Our results underscore the importance of systematic, multiwavelength counterpart searches for EP-discovered transients, which are essential for establishing the nature of individual events and building a well-characterized eFXT population. While EP/WXT enables wide-field discovery, EP/FXT—despite its smaller field of view ($\sim1\ \mathrm{deg}^{2}$)—probes an observable volume of $\sim10\%$ that of WXT owing to its higher sensitivity, and provides superior localization accuracy ($\sim10^{\prime\prime}$) and spectral quality. Real-time counterpart searches and prompt multiwavelength follow-up observations are critical for fully exploiting the scientific potential of EP.

\section*{Acknowledgments}
%\begin{acknowledgments}

This work is based on data obtained with the Einstein Probe (EP), a space mission supported by the Strategic Priority Program on Space Science of the Chinese Academy of Sciences (CAS), in collaboration with the European Space Agency (ESA), the Max-Planck-Institute for Extraterrestrial Physics (MPE, Germany), and the Centre National d’Études Spatiales (CNES, France).

W.X.L. is supported by the National Natural Science Foundation of China (NSFC; grants 12120101003, 12233008, and 12373010), the National Key R\&D Program of China (2022YFA1602902 and 2023YFA1607804), and the Strategic Priority Research Program of CAS (XDB0550100 and XDB0550000). L.Z.W. is supported by NSFC grant 12573050 and in part by CAS through a grant to the Chinese Academy of Sciences South America Center for Astronomy (CASSACA). L.Z.W., W.X.L., N.C.S., E.T., J.S.H., F.E.B., and J.C. are supported by the CASSACA Key Research Project (E52H540301).

M.N. acknowledges support from the European Research Council (ERC) under the European Union's Horizon 2020 research and innovation programme (grant agreement No. 948381). A.A. and T.-W.C. acknowledge support from the Ministry of Education Yushan Fellow Program (MOE-111-YSFMS-0008-001-P1) and the National Science and Technology Council of Taiwan (NSTC 114-2112-M-008-021-MY3). F.E.B. and E.T. acknowledge funding from ANID through CATA BASAL (FB210003) and FONDECYT Regular grants 1241005 and 1250821.

H.S. acknowledges support from NSFC (grant 12573049) and the Young Elite Scientists Sponsorship Program by the China Association for Science and Technology (YESS20240218). S.Y. is supported by NSFC grant 12303046, the Startup Research Fund of the Henan Academy of Sciences (242041217), and the Joint Fund of the Henan Province Science and Technology R\&D Program (235200810057).

Part of the funding for GROND (both hardware and personnel) was provided by the Leibniz Prize awarded to Prof. G. Hasinger (DFG grant HA 1850/28-1). The VISTA Hemisphere Survey data products served at Astro Data Lab are based on observations collected at the European Organisation for Astronomical Research in the Southern Hemisphere (ESO) under programme 179.A-2010, and/or data products derived from these observations. This work is based on observations made with the Thai Robotic Telescope under program ID TRTToO~2024002, which is operated by the National Astronomical Research Institute of Thailand (Public Organization).

\vspace{5mm}
\facilities{Einstein Probe (EP), Swift (XRT and UVOT), Fermi-GBM, VLT, Max Planck:2.2m (GROND)}


\software{astropy \citep{2013A&A...558A..33A},  
          Numpy \citep{2020Natur.585..357H},
          Matplotlib \citep{2007CSE.....9...90H}
          }

\appendix

\section{Details of Candidates}
\label{appendix:candidates}

Here we provide details of the discovery and identification of the matched candidates listed in Tables~\ref{table:candidates_highconf},~\ref{table:candidates_unverified}.

\subsection{EP240315a}

EP240315a was discovered by EP/WXT at 20:10:44~UTC on 15 March 2024, with a duration of $T_{90,X}$= 1034\,s. The gamma-ray counterpart,  GRB 240315C, was subsequently detected by \textit{Swift}/BAT and \textit{Konus}-Wind 372\,s after the initial WXT trigger. The optical counterpart, AT2024eju, was identified by ATLAS approximately 1.1\,hours later. Spectroscopy established a redshift of $z=4.859$, indicating that EP/WXT captured the soft X-ray prompt emission of a high-redshift GRB \citep{liu2025soft,levan2024fast,2024ApJ...969L..14G}.

\subsection{EP240414a}

EP240414a was discovered by EP/WXT at 09:49:10 UTC on 14 April 2024, and no contemporaneous gamma-ray signal was detected. The optical counterpart, SN~2024gsa, was subsequently identified in the luminous host galaxy SDSS~J124601.99$-$094309.3 and classified as  (Ic-BL) SN. The light curve of SN~2024gsa exhibits a distinctive multi-bump evolution. The observed soft X-ray spectrum and luminosity at $z = 0.401$ suggest that EP240414a originated from the interaction of a weak relativistic jet with an extended envelope surrounding the progenitor star~\citep{sun2025fast,van2025einstein,hamidani2025ep240414a}.

\subsection{EP240425a}

EP240425a was discovered by EP/WXT at 20:50:33 UTC on 25 April 2024, with no prompt gamma-ray signal detected. A rapidly fading optical counterpart was subsequently identified, while the radio emission continued to brighten for $\gtrsim170$ days. In the absence of a reliable redshift measurement, the physical origin of the transient cannot be established definitively. Nevertheless, the observed multi-wavelength behavior is broadly consistent with expectations for a jetted tidal disruption event (Zhao et al. in prep).

\subsection{EP240506a}

EP240506a was triggered via the VHF EP/WXT alert channel at 05:01:39 UTC on 6 May 2024 during the commissioning phase of EP. However, as it did not activate the BeiDou alert system, no automated follow-up observations were initiated, resulting in a missed decay phase of the transient. The transient lasted for approximately 50\,s, reaching a peak flux of $1\times10^{-8}\ \mathrm{erg\cdot s^{-1}\cdot cm^{-2}}$~\citep{li2024ep240506a}. No counterpart was identified in the immediate follow-up observations~\citep{2024GCN.36419....1J,2024GCN.36417....1P,2024GCN.36414....1C,2024GCN.36413....1T,2024GCN.36412....1P,2024GCN.36408....1A}. In this work, we find that the event is probably associated with an SN (AT~2024ofs). %that emerged several weeks after the X-ray transient, suggesting a delayed optical counterpart. 
Spectroscopic observation of its host galaxy yields a redshift of $z = 0.12$. %, consistent with the possible origin of a weak relativistic jet. 
Further details of EP240506a are discussed in Section~\ref{subsec:Ep240506a}.

\subsection{EP241030a}

EP241030a was detected by EP/WXT at 06:33:18 UTC on 30 October 2024, with a duration of approximately 50\,s. 
The event was found to be spatially and temporally coincident with GRB~241030A, and was subsequently identified as its X-ray afterglow counterpart~\citep{2024GCN.37997....1W}. 
The prompt gamma-ray emission was observed earlier by \textit{Fermi}-GBM and \textit{Swift}-BAT at 05:48:03~UTC on the same day, about 45 minutes before the EP detection~\citep{fermi2024grb,klingler2024grb}. 
A bright optical counterpart was also detected by \textit{Swift}-UVOT, and follow-up spectroscopic observations with Keck/LRIS measured a redshift of $z=1.411$~\citep{zheng2024grb}.

\subsection{EP250108a}

EP250108a was discovered by EP/WXT at 12:30:28.34 UTC on 8 January 2025, with no significant gamma-ray counterpart detected~\citep{li2025ep250108a}. The optical counterpart, SN~2025kg, %was initially flagged as FBOT through 
was identified through follow-up observations, although no associated radio emission was observed~\citep{eyles2025ep250108a,zhu2025ep250108a,izzo2025ep250108a,zou2025ep250108a,schroeder2025ep250108a,eyles2025ep250108ahop,an2025ep250108a}. Subsequent spectroscopic analysis classified SN~2025kg as a type Ic-BL SN, originating from the core collapse of a Wolf-Rayet star at redshift $z = 0.176$~\citep{zhu2025ep250108a,xu2025ep250108a,li2025extremely}. Further investigations suggest that the observed X-ray emission likely arose from the interaction between a mildly relativistic outflow and the surrounding circumstellar material~\citep{li2025extremely,srinivasaragavan2025ep250108a,zhu2025,zhu2025ep250108a}.

\subsection{EP250226a}

EP250226a was triggered by EP/WXT at 06:35:16 UTC on 26 February 2025, followed by automatic FXT observations. 
The event was associated with GRB~250226A, which had been detected by \textit{Fermi}-GBM, 
Konus-Wind, \textit{Swift}-BAT and GECAM-B. The EP X-ray detection occurred about 20\,s after the \textit{Fermi}-GBM trigger~\citep{fermi2025grb250226a,jiang2025ep250226a}. 
An optical counterpart was subsequently identified using the VLT, and spectroscopic observations determined a redshift of $z=3.315$~\citep{zhu2025ep250226a}.

\subsection{EP250304a} 

EP250304a was discovered by EP/WXT at 01:32:30 UTC on 4 March 2025, with a duration of approximately 1200\,s~\citep{Chen250304a,zhang2025ep250304a}. 
An optical counterpart was detected about 1.12\,hr after the WXT trigger~\citep{liu2025ep250304a}. 
Subsequent spectroscopic observations with the VLT measured a redshift of $z=0.2$ and classified the transient as a type Ic-BL SN~\citep{saccardi2025ep250304a,izzo2025ep250304a}. 
This makes EP250304a the third SN associated with an eFXT detected by EP.

\subsection{EP250427a}

EP250427a was detected by EP/WXT at 03:38:45 UTC on 27 April 2025. The true onset time ($T_0$) remains unconstrained, as the satellite was within the SAA region prior to the trigger. The event is associated with a sub-threshold burst, GRB 250427A, detected by Fermi-GBM~\citep{wang2025ep250427a,ravasio2025grb}. An optical counterpart was detected 4 hr after trigger~\citep{perez2025ep,liu2025ep250427a}. Spectroscopic observations conducted with by Keck/LRIS and VLT/X-Shooter confirmed a redshift of $z=1.52$~\citep{chornock2025ep250427a,saccardi2025ep250427a}. 

\subsection{EPW20240326aa}

EPW20240326aa was initially flagged as a transient due to a prominent flare lasting approximately $10^{4}\ \mathrm{s}$, accompanied by apparent periodic oscillations. However, this source was later determined to be spurious, likely caused by background fluctuations, as similar patterns were observed across other CMOS detectors. Consequently, it was reclassified as an \textit{unverified sources} attributed to the potential of instrumental background variations.

\subsection{WXT~J160346+19355}\label{J160346}

WXT~J160346+19355 was identified as an \textit{unverified source}, exhibiting a relatively steady X-ray flux over several days, followed by non-detections. There is a historical X-ray source 2CXO J160346.6+193540 in Chandra Source Catalog (CSC). The position of the corresponding optical object, as reported by Gaia and suggested to be a brightening cataclysmic variable (CV), is spatially consistent with the known CV CRTS~CSS160906~J160346+193540 and 2CXO J160346.6+193540. We conclude that the X-ray emission and the optical counterpart might originate from the same source. The detections on those days were made possible by ultra-long effective exposures of $\sim 2 \times 10^{4}\ \mathrm{s}$, accumulated during the instrument calibration phase.

\subsection{Others}

Based on the large position uncertainties of WXT and low signal-to-noise ratios, the remaining \textit{unverified sources} are likely attributable to chance spatial and temporal coincidences, given the marginal significance of their X-ray detections and the absence of follow-up observations or any corroborating evidence supporting a physical association. 

\section{Multi-wavlength data of EP240506a/AT~2024ofs}\label{appendix:data}

The UV/optical/IR photometric measurements of EP240506a/AT~2024ofs are summarized in Table~\ref{tab:data}.

\startlongtable
\begin{deluxetable*}{cccc}
\tablecaption{Multi-wavelength photometry of EP240506a/AT~2024ofs. \label{tab:data}}
\tablehead{
\colhead{Phase (days)} & \colhead{Telescope} & \colhead{Band} & 
\colhead{Mag}
}
\startdata
1.203 & TRT & \textit{I} & $>21.0$ \\
1.232 & TRT & \textit{R} & $>21.6$ \\
1.428 & XLT & \textit{g} & $>23.10$ \\
1.909 & Swift/UVOT & uvw2 & $>21.77$ \\
1.909 & Swift/UVOT & uvm2 & $22.57 \pm 0.43$ \\
1.909 & Swift/UVOT & \textit{u} & $>20.29$ \\
1.909 & Swift/UVOT & \textit{v} & $>18.66$ \\
1.909 & Swift/UVOT & \textit{b} & $>19.49$ \\
1.909 & Swift/UVOT & uvw1 & $>20.93$ \\
2.704 & ATLAS & \textit{c} & $>\textit{19.68}$ \\
2.707 & ATLAS & \textit{c} & $>\textit{19.69}$ \\
2.724 & ATLAS & \textit{c} & $>\textit{19.66}$ \\
3.004 & GROND\tablenotemark{a} & \textit{H} & $>18.3$ \\
3.004 & GROND\tablenotemark{a} & \textit{J} & $>19.1$ \\
3.009 & GROND\tablenotemark{a} & \textit{K} & $>17.3$ \\
3.698 & ATLAS & \textit{c} & $>19.71$ \\
3.702 & ATLAS & \textit{c} & $>19.70$ \\
3.709 & ATLAS & \textit{c} & $>19.68$ \\
3.711 & ATLAS & \textit{c} & $>19.65$ \\
3.720 & ATLAS & \textit{c} & $>19.69$ \\
3.721 & ATLAS & \textit{c} & $>19.60$ \\
3.756 & ATLAS & \textit{c} & $>19.57$ \\
4.922 & ATLAS & \textit{o} & $>18.01$ \\
4.927 & ATLAS & \textit{o} & $>17.90$ \\
4.930 & ATLAS & \textit{o} & $>17.75$ \\
5.948 & ATLAS & \textit{o} & $>19.39$ \\
5.955 & ATLAS & \textit{o} & $>19.34$ \\
5.968 & ATLAS & \textit{o} & $>19.31$ \\
6.712 & ATLAS & \textit{c} & $>19.47$ \\
7.720 & ATLAS & \textit{c} & $20.69 \pm 0.34$ \\
11.675 & ATLAS & \textit{o} & $>19.05$ \\
11.678 & ATLAS & \textit{o} & $>19.15$ \\
12.951 & ATLAS & \textit{o} & $>18.84$ \\
19.086 & ZTF & \textit{r} & $19.84 \pm 0.36$ \\
19.726 & ATLAS & \textit{o} & $>18.73$ \\
19.734 & ATLAS & \textit{o} & $>18.67$ \\
19.750 & ATLAS & \textit{o} & $>18.66$ \\
20.979 & ATLAS & \textit{o} & $>19.02$ \\
20.983 & ATLAS & \textit{o} & $>19.05$ \\
21.064 & ZTF & \textit{g} & $19.78 \pm 0.12$ \\
21.133 & ZTF & \textit{r} & $20.27 \pm 0.21$ \\
21.981 & ATLAS & \textit{o} & $20.27 \pm 0.30$ \\
23.084 & ZTF & \textit{g} & \(20.03 \pm 0.12\) \\
23.127 & ZTF & \textit{r} & \(20.34 \pm 0.18\) \\
24.118 & Pan-STARRS & \textit{w} & \(20.26 \pm 0.04\) \\
24.129 & Pan-STARRS & \textit{w} & \(20.19 \pm 0.05\) \\
24.140 & Pan-STARRS & \textit{w} & \(20.17 \pm 0.06\) \\
24.152 & Pan-STARRS & \textit{w} & \(20.24 \pm 0.06\) \\
24.953 & ATLAS & \textit{c} & \(20.44 \pm 0.32\) \\
25.043 & ZTF & \textit{g} & \(20.37 \pm 0.17\) \\
25.928 & ATLAS & \textit{c} & \(20.08 \pm 0.18\) \\
27.024 & ZTF & \textit{r} & \(20.84 \pm 0.22\) \\
27.065 & ZTF & \textit{g} & \(21.07 \pm 0.31\) \\
29.123 & Pan-STARRS & \textit{w} & \(21.09 \pm 0.23\) \\
29.133 & Pan-STARRS & \textit{w} & \(21.11 \pm 0.23\) \\
29.146 & Pan-STARRS & \textit{w} & \(20.93 \pm 0.22\) \\
29.157 & Pan-STARRS & \textit{w} & \(20.95 \pm 0.17\) \\
30.129 & ATLAS & \textit{o} & \(20.58 \pm 0.34\) \\
36.097 & Pan-STARRS & \textit{w} & \(21.83 \pm 0.15\) \\
\enddata
\tablenotetext{a}{GROND magnitudes are obtained in the Vega system and calibrated using REFCAT2 sources in the field of view.}
\end{deluxetable*}

\section{Model Fitting of AT~2024ofs}\label{appendix:fitting}

\begin{figure*}[htb!]
    \centering
    \includegraphics[width=0.8\linewidth]{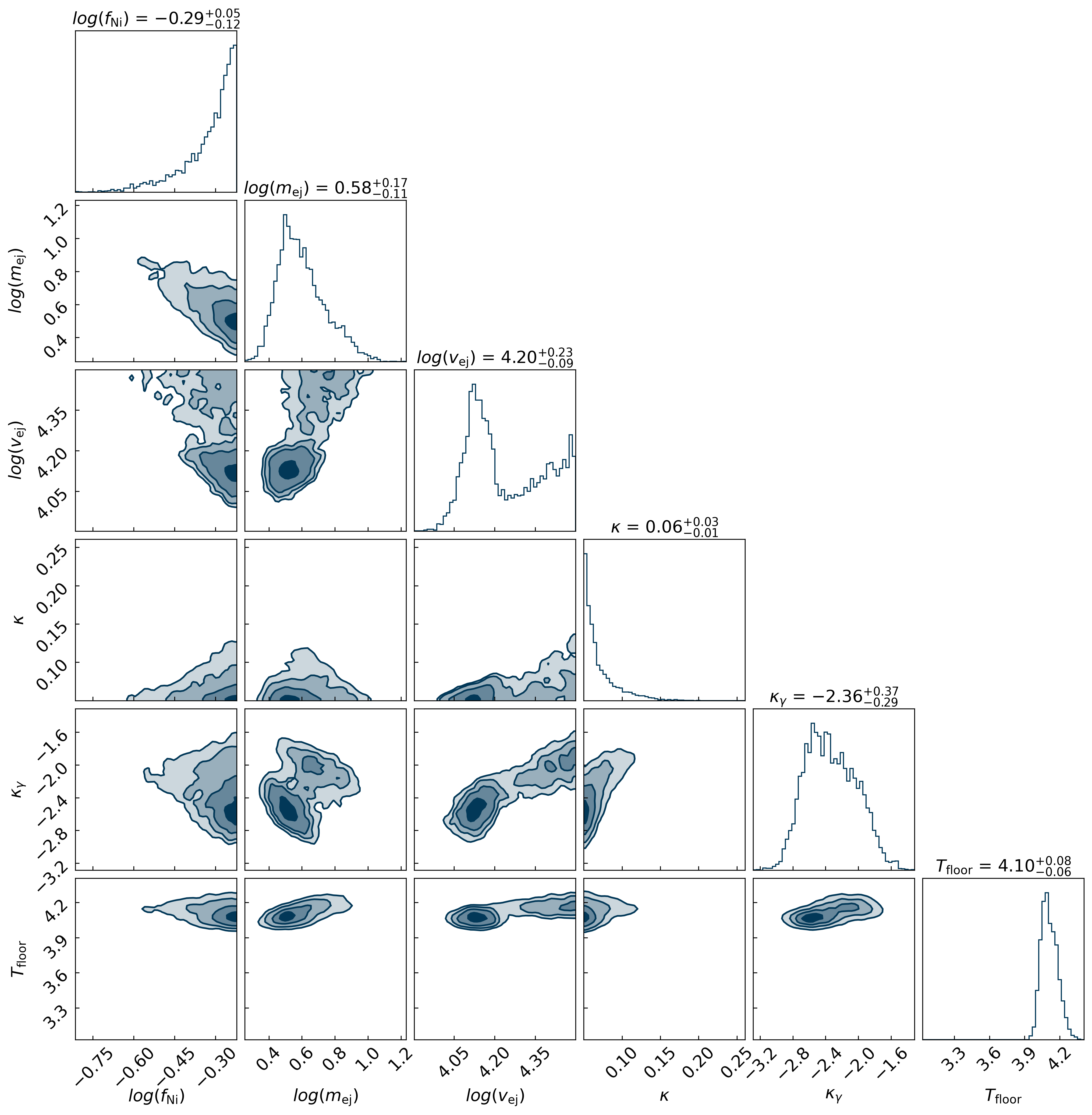}
    \caption{Corner plot of the posterior distributions from the Arnett-model light-curve fitting of AT 2024ofs. The notation log denotes the base-10 logarithm. The free parameters are: $\log(f_{\rm Ni})$ — the mass fraction of radioactive $^{56}$Ni, $\log(m_{\rm ej})$ — the ejecta mass, $\log(v_{\rm ej})$ — the characteristic ejecta velocity, $\kappa$ — the grey opacity of the ejecta, $\kappa_{\gamma}$ — the gamma-ray opacity, and $T_{\rm floor}$ — the temperature floor.}
    \label{fig:corner}
\end{figure*}

We present the posterior distributions obtained from the Bayesian inference of the AT~2024ofs  light curve using the Arnett model implenmented in \texttt{Redback}, as shown in Figure~\ref{fig:corner}. The inferred parameters--including $f_{\mathrm{Ni}}$, $v_{\mathrm{ej}}$, and $T_{\mathrm{floor}}$--exhibit strong degeneracies and tend to drift toward the boundaries or unphysical regions of the parameter space.

\bibliography{sample631}{}
\bibliographystyle{aasjournal}



\end{document}